\documentclass[aps,pre,superscriptaddress,amsmath,amssymb,twocolumn]{revtex4}
\usepackage{graphicx}
\usepackage{hyperref}
\usepackage{color}

\definecolor{darkblue}{rgb}{0.0,0.0,0.5}
\hypersetup{colorlinks,breaklinks,
            linkcolor=darkblue,urlcolor=darkblue,
            anchorcolor=darkblue,citecolor=darkblue}

\date{\today}

\begin{document}

\title{Packing and Self-assembly of Truncated Triangular Bipyramids}

\author{Amir Haji-Akbari}
%\email{hajakbar@princeton.edu}
\affiliation{Department of Chemical Engineering, University of Michigan, Ann Arbor MI 48109}
\affiliation{Department of Chemical and Biological Engineering, Princeton University, Princeton NJ 08540}

\author{Elizabeth R. Chen}
%\email{bethchen@seas.harvard.edu}
\affiliation{Department of Chemical Engineering, University of Michigan, Ann Arbor MI 48109}
\affiliation{Harvard School of Engineering and Applied Sciences, Harvard University, Cambridge MA 02138}

\author{Michael Engel}
%\email{engelmm@umich.edu}
\affiliation{Department of Chemical Engineering, University of Michigan, Ann Arbor MI 48109}

\author{Sharon C. Glotzer}
\email{sglotzer@umich.edu}
\affiliation{Department of Chemical Engineering, University of Michigan, Ann Arbor MI 48109}
\affiliation{Department of Materials Science and Engineering, University of Michigan, Ann Arbor MI 48109}

\date{\today}

\begin{abstract}
Motivated by breakthroughs in the synthesis of faceted nano- and colloidal particles, as well as theoretical and computational studies of their packings, we investigate a family of truncated triangular bipyramids. We report dense periodic packings with small unit cells that were obtained via numerical and analytical optimization. The maximal packing fraction $\phi_{\max}$ changes continuously with the truncation parameter $t$. Eight distinct packings are identified based on discontinuities in the first and second derivatives of $\phi_{\max}(t)$. These packings differ in the number of particles in the fundamental domain (unit cell) and the type of contacts between the particles. In particular, we report two packings with four particles in the unit cell for which both $\phi_{\max}(t)$ and $\phi'_{\max}(t)$ are continuous and the discontinuity occurs in the second derivative only. In the self-assembly simulations that we perform for larger boxes with 2048 particles, only one out of eight packings is found to assemble.
In addition, the degenerate quasicrystal reported previously for triangular bipyramids without truncation [\href{http://prl.aps.org/abstract/PRL/v107/i21/e215702}{Haji-Akbari~\emph{et al.}, \emph{Phys. Rev. Lett.} \textbf{107}: 215702 (2011)}]  assembles for truncations as high as $0.45$. The self-assembly propensities for the  structures formed in the thermodynamic limit are explained using the isoperimetric quotient of the particles and the coordination number in the disordered fluid and in the assembled structure.
\end{abstract}

\maketitle

\section{Introduction\label{section:intro}}

Hard core interactions are idealized representations of excluded volume effects that are prevalent in dense states of matter such as liquids and solids, and have been shown to approximate short-range repulsions~\cite{WeeksChandlerAndersonPRA1971}. In atomic and molecular systems, such repulsions arise when atoms and molecules get close enough for their electronic shells to overlap and repel. It is therefore no surprise that hard spheres have been successful models of dense liquids and solids for decades~\cite{AlderWainwright1957, WoodJacobson1957, AlderGassWainright1970, AndersonChandler1972, WoodcockJChemSocFaraday1976,TarazonaPRA1985,LevesquePRE1994,ZamponiJCP2005}. At larger length scales, hard core interactions are used to predict the thermodynamic behavior of concentrated colloidal suspensions since colloidal particles can, under certain circumstances, behave like hard particles. For example, long-range electrostatic repulsions between charged nanoparticles are screened by ionic solutions, effectively becoming short-range repulsions~\cite{HansenLowenAnnRevPhysChem2000}, which can be approximated with hard core interactions~\cite{WeeksChandlerAndersonPRA1971}. The phases expected for such colloidal systems can be predicted through theoretical and computational studies of the corresponding hard particle system. Recent breakthroughs in the synthesis of colloidal particles of various shapes~\cite{Ahmadi1996, JinMirkinNature2003,  MurphyJACS2004, ChampoinPolymerPNAS2007, MirkinAngewChemIntEd2009, MirkinJACS2010, MurrayACSNano2012, MurrayJACS2012, GeisslerYang2012,AstrucCCR2012, MirkinJACS2012, MirkinScience2012,MirkinJACS2013} have spurred an increased interest in studies of hard particle systems.

A remarkable fact about hard particles is their ability to spontaneously form ordered phases at sufficiently high packing fractions. While the emergent order is typically periodic in the form of a crystal, quasiperiodic and degenerate~\footnote{In some cases, (quasi-)periodic order can be replaced by degenerate order in the densest packing state~\cite{HajiAkbariDQC2011,WojciechowskiPRL1991}. Despite their random character, degenerate (quasi)crystals have long-range order.} order are also possible. These phases are stabilized by entropy alone and are observed in computational studies of various hard particle shapes including spheres~\cite{AlderWainwright1957,WoodJacobson1957}, spherocylinders~\cite{BolhuisFrenkel1997}, thin disks~\cite{FrenkelEppengaPRL1982, DijkstraPRL2012}, ellipsoids~\cite{FrenkelPRL1984, DonevScience2004, DonevStillingerPRL2004, schillingPRE2007, RaduSchilling2009}, dumbbells~\cite{DijkstraPRE2013}, tetrahedra~\cite{HajiAkbariEtAl2009, HajiAkbaricondmat2011}, triangular bipyramids~\cite{HajiAkbariDQC2011}, superballs, cubes and octahedra~\cite{Dijkstra2012,DijkstraPNAS2012}, snowman particles~\cite{DijkstraSnowman2012}, squares~\cite{EscobedoSoftMatter2012}, space-filling polyhedra~\cite{JohnEscobedoJPCB2005, JohnEscobedo2008,EscobedoNatureMaterials2011} and many other polyhedra~\cite{PabloScience2012}. The preponderance of optical, electrical, magnetic and mechanical properties of ordered structures formed from such particles~\cite{AlivisatosScience1996,PengAdvMatSemiconductor2003} makes knowledge and prediction of their expected thermodynamic assemblies of particular current interest.

The feasibility of disorder-order transitions can be understood from a packing perspective as all permissible configurations of hard particle systems are packings of the corresponding shape. More particularly, the structure that is thermodynamically stable in the limit of infinite pressure is the packing (or, if there are several equally dense packings, is among the packings) with the largest possible packing fraction. This is seen from the Gibbs free energy $G(P,T)=PV-ST$ of the hard particle system. As the entropy is bounded from below and decreases with increasing pressure, the Gibbs free energy is dominated by the $PV$ term as $P\rightarrow\infty$. The structure with the smallest volume (largest packing fraction) will eventually be stable. Solving the packing problem is therefore equivalent to identifying the equilibrium structure of the corresponding hard particle system in the infinite-pressure limit. All known densest packings of two- and three-dimensional objects are ordered~\cite{BezdekConvexPacking2010}, which necessitates a disorder-order transition at some finite pressure and density. However, it is also known that the ordered structures formed by many hard particle systems at intermediate densities can be geometrically unrelated to their corresponding densest packings~\cite{HajiAkbaricondmat2011, HajiAkbariDQC2011, PabloACSNanot2012, PabloScience2012}.

One hard particle with a rich and unconventional phase behavior is the hard tetrahedron. The densest known packing of tetrahedra is a double-dimer lattice with four particles in its fundamental domain forming two pairs or `dimers'~\cite{Kallus2010, TorquatoJiaoarxiv, Chen2010}. At intermediate densities, however, tetrahedra self-assemble into a dodecagonal quasicrystal~\cite{HajiAkbariEtAl2009}. The $(3.4.3^2\!.4)$ approximant of the quasicrystal is only slightly less dense than the double-dimer lattice and is more stable at intermediate densities~\cite{HajiAkbaricondmat2011}.  A similar trend is observed for the closely related hard triangular bipyramid, which is a dimer of tetrahedra. Triangular bipyramids form a degenerate dodecagonal quasicrystal, a structure that is identical to the tetrahedron-based quasicrystal on the monomer level but random in the pairing of constituent tetrahedra into dimers in the nearest neighbor network. At intermediate densities, both the degenerate quasicrystal and its approximant are again more stable than the densest packing of triangular bipyramids, which is structurally identical to the double-dimer lattice of tetrahedra~\cite{HajiAkbariDQC2011}.

Damasceno~\emph{et~al.} studied the self assembly and densest packing behavior of  a family of truncated tetrahedra~\cite{PabloACSNanot2012}. The building blocks were obtained by truncating the vertices of a regular tetrahedron. The particle geometry ranged from a perfect tetrahedron with truncation parameter $t=0$ to a perfect octahedron with $t=1$. It was shown that the formation of the quasicrystal for the tetrahedron at intermediate packing fractions is robust and occurs for truncations up to $t\le0.45$. Several unexpected structures self-assemble at higher truncations, including diamond, $\beta$-tin, and a crystal isostructural to high-pressure lithium. A previously unreported space-filling polyhedron was obtained for $t=1/2$. Most notably, the assembled structures of truncated tetrahedra are typically distinct from the densest packings obtained in numerical compressions of small unit cells~\cite{PabloACSNanot2012}. In addition to the potentially complex phase behavior, studying truncated polyhedra is relevant from a practical perspective. Experimentally synthesized nano- and colloidal particles are not perfect, but typically have various types of imperfections~\cite{DamascenoImperfect}, one of which can be truncated vertices~\cite{MirkinJACS2010}. Studying the role of truncation allows us to investigate the robustness of the self-assembly process to such imperfections. 

Motivated by these considerations, we use techniques similar to those used in Refs.~\cite{HajiAkbariEtAl2009,Chen2010,HajiAkbariDQC2011,PabloACSNanot2012} to investigate the packing and self-assembly of hard truncated triangular bipyramids. We study the densest packings with small repeating unit cells using Monte Carlo simulations and further analytical and numerical optimization. Eight distinct packing types, six with two particles in the unit cell and two with four particles in the unit cell, are identified for different values of truncation based on discontinuities in the first and the second derivatives of $\phi_{\max}(t)$. The two four-particle packings are observed for a narrow range of truncations and are separated by a discontinuity in $\phi''_{\max}(t)$. We also simulate large systems of truncated triangular bipyramids and observe that only one out of eight packing types forms in simulation. We observe that the quasicrystal formation is robust and occurs for truncations as high as $0.45$. Finally, we explain the self-assembly potential of different building blocks by their isoperimetric quotient and the coordination number of their disordered fluid and find that only the structures that are locally similar to their disordered fluid tend to form in the self-assembly simulations of the bulk system. 

This paper is organized as follows. Geometrical notations and definitions are given in Section~\ref{subsec:notation}. Technical details of Monte Carlo simulations are presented in Section~\ref{subsec:MC}. Section~\ref{subsec:packing} outlines our formulation of the constrained packing problem and the approach used for its solution. The solutions of the packing problem for truncated TBPs are thoroughly discussed in Section~\ref{subsec:densePackings}. Section~\ref{subsec:self_assembly} discusses the self-assembly simulations while the role of the local structure and the particle shape on self-assembly are discussed in Section~\ref{subsec:shapeLocalOrder}. And Section~\ref{sec:conclusion} is reserved for discussions and concluding remarks.

\section{Methods\label{section:methods}}
\subsection{Geometric notations and definitions\label{subsec:notation}}
Let $\mathcal{P}$ be a perfect (non-truncated) triangular bipyramid (TBP) centered at the origin. It is the convex hull of its five vertices
\begin{subequations}
\label{eq:TBP_vertices}
\begin{align}
\textbf{o} &= \left(+2,+2,+2\right),\\
\textbf{p} &= \left(+2,-1,-1\right),\\
\textbf{q} &= \left(-1,+2,-1\right),\\
\textbf{r} &= \left(-1,-1,+2\right),\\
\textbf{s} &= \left(-2,-2,-2\right).
\end{align}
\end{subequations}
To truncate the TBP, we replace these vertices by new vertices positioned along the edges as specified by a truncation parameter $t$. Two new vertices are positioned uniformly and symmetrically on each edge in-between the old vertices, $t=0$, and the edge mid centers, $t=1$. The truncated triangular bipyramid (tTBP) $\mathcal{P}_t$ is then defined as the convex hull of the following 18 vertices
\begin{subequations}
	\label{eq:tTBPvertices}
	\begin{align}
		\textbf{op}=(1-\tfrac{t}2)\textbf{o}+\tfrac{t}2\textbf{p}, &~~~~& \textbf{po}=(1-\tfrac{t}2)\textbf{p}+\tfrac{t}2\textbf{o},
		\\
		\textbf{oq}=(1-\tfrac{t}2)\textbf{o}+\tfrac{t}2\textbf{q}, &~~~~& \textbf{qo}=(1-\tfrac{t}2)\textbf{q}+\tfrac{t}2\textbf{o},
		\\
		\textbf{or}=(1-\tfrac{t}2)\textbf{o}+\tfrac{t}2\textbf{r}, &~~~~& \textbf{ro}=(1-\tfrac{t}2)\textbf{r}+\tfrac{t}2\textbf{o},
		\\
		\textbf{sp}=(1-\tfrac{t}2)\textbf{s}+\tfrac{t}2\textbf{p}, &~~~~& \textbf{ps}=(1-\tfrac{t}2)\textbf{p}+\tfrac{t}2\textbf{s},
		\\
		\textbf{sq}=(1-\tfrac{t}2)\textbf{s}+\tfrac{t}2\textbf{q}, &~~~~& \textbf{qs}=(1-\tfrac{t}2)\textbf{q}+\tfrac{t}2\textbf{s},
		\\
		\textbf{sr}=(1-\tfrac{t}2)\textbf{s}+\tfrac{t}2\textbf{r}, &~~~~& \textbf{rs}=(1-\tfrac{t}2)\textbf{r}+\tfrac{t}2\textbf{s},
		\\
		\textbf{qr}=(1-\tfrac{t}2)\textbf{q}+\tfrac{t}2\textbf{r}, &~~~~& \textbf{rq}=(1-\tfrac{t}2)\textbf{r}+\tfrac{t}2\textbf{q},
		\\
		\textbf{rp}=(1-\tfrac{t}2)\textbf{r}+\tfrac{t}2\textbf{p}, &~~~~& \textbf{pr}=(1-\tfrac{t}2)\textbf{p}+\tfrac{t}2\textbf{r},
		\\
		\textbf{pq}=(1-\tfrac{t}2)\textbf{p}+\tfrac{t}2\textbf{q}, &~~~~& \textbf{qp}=(1-\tfrac{t}2)\textbf{q}+\tfrac{t}2\textbf{p}.
	\end{align}
\end{subequations}
We denote a vertex of $\mathcal{P}_t$ by $V[\textbf{x}]$, where $\textbf{x}$ is any of the points given in Eq.~(\ref{eq:tTBPvertices}). An edge is denoted either by its endpoints or by any two points that lie along the line segment connecting its endpoints. Similarly, a face is denoted by any three points that are coplanar with the face. For instance, an edge with the endpoints $\textbf{op}$ and $\textbf{po}$ is denoted by $E[\textbf{o},\textbf{p}]$ and the hexagonal face containing the points $\textbf{op}, \textbf{po}, \textbf{pq}, \textbf{qp}, \textbf{qo}$, and $\textbf{oq}$ by $F[\textbf{o},\textbf{p},\textbf{q}]$. 

\begin{figure}
	\begin{center}
		\includegraphics[width=\columnwidth]{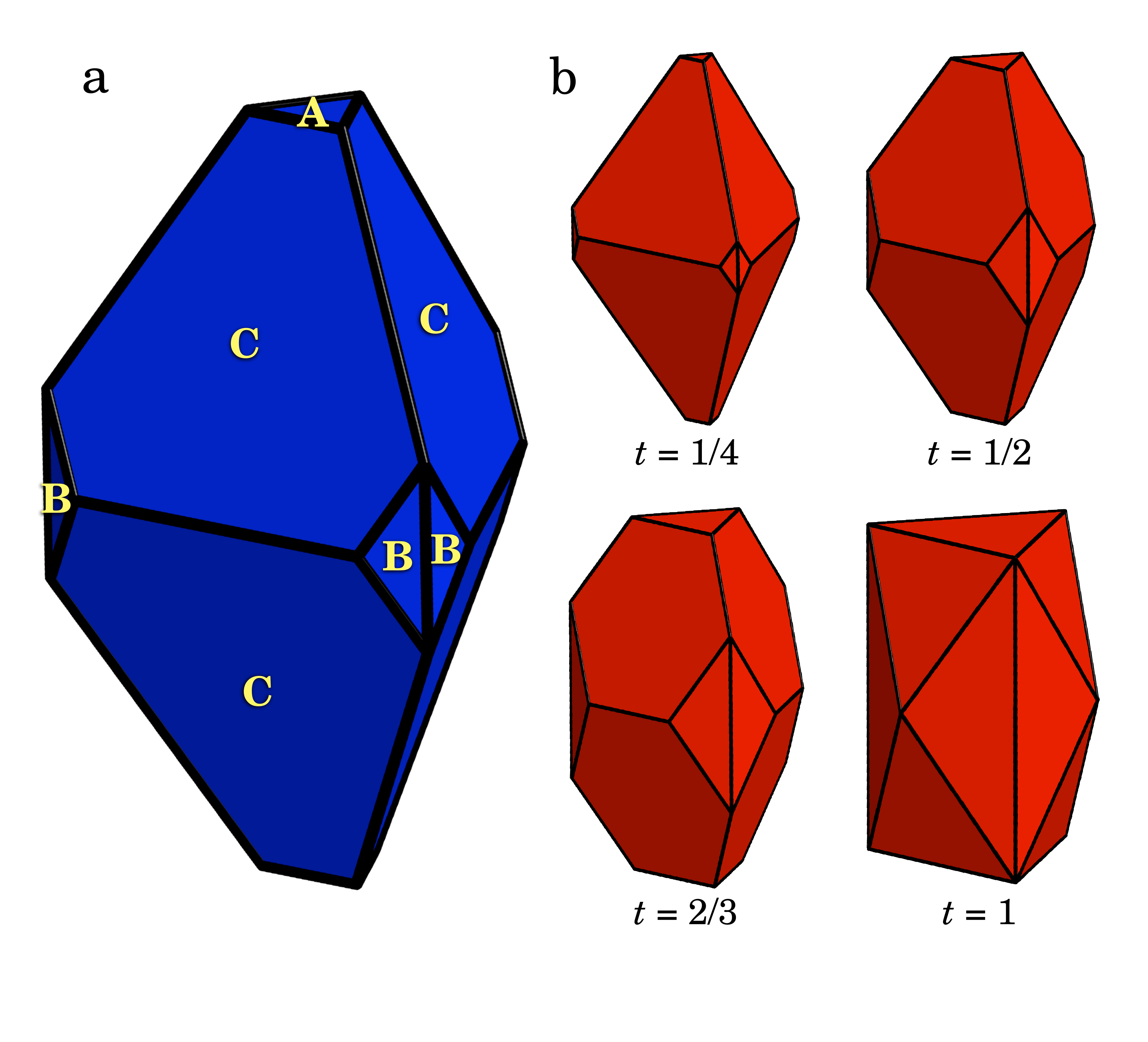}
		\caption{(color online) (a)~A truncated triangular bipyramid consists of two regular tetrahedra joined together at one face, truncated symmetrically at the five vertices. Polar triangles, equatorial triangles, and peripheral hexagons are denoted by A, B, and C, respectively.  (b)~Truncated triangular bipyramids for different values of the truncation parameter $t${\label{fig:TBP:tTBP}. 
		}}
	\end{center}
\end{figure}

Every tTBP has three types of faces that are distinguished in Fig.~\ref{fig:TBP:tTBP}a. The two triangular faces perpendicular to the three-fold axis of the particle are referred to as \emph{polar triangles} and are labeled A in the figure. The six triangles with an edge parallel to the three-fold axis are called \emph{equatorial triangles} and are labeled B. The remaining six hexagonal faces are referred to as \emph{peripheral hexagons} and are labelled C.

The volume and the surface area of $\mathcal{P}_t$ are given by
\begin{eqnarray}
V(t) &=& V_p\left(1-\tfrac{3}{8}t^3\right),\\
S(t) &=& S_p\left[1-\tfrac16(6-\sqrt6)t^2\right],
\end{eqnarray}
where $V_p=18$ and $S_p=27\sqrt3$ are the volume and the surface area of $\mathcal{P}$, respectively. Fig.~\ref{fig:TBP:tTBP}b depicts tTBPs for a few special values of $t$. 

\subsection{Monte Carlo simulations\label{subsec:MC}}
We perform Monte Carlo (MC) simulations of tTBPs with truncations ranging from zero to unity. The edge length of a perfect non-truncated TBP is chosen as the length scale and pressure is measured as $P^*=PV(t)/k_BT$. Overlaps between tTBPs are detected using the Gilbert-Johnson-Keerthi (GJK) algorithm~\cite{GilbertGJK1988}. Each MC cycle is comprised of one trial move per particle, which is either a trial translation or a trial rotation with equal probabilities. All simulations are performed in boxes with periodic boundary conditions. 

Two types of MC simulations are carried out in this study. In order to identify dense small-cell packings of tTBPs, isotension MC simulations are performed in small unit cells, i.e.~in boxes containing up to eight particles. Such simulations include an additional trial move per cycle wherein both the size and the shape of the simulation box are varied. The box move is accepted or rejected according to the conventional Metropolis criterion in the NPT ensemble. All simulations with small boxes start at a low pressure of $P^*=1$. The pressure is then exponentially increased every $50\,000$ MC cycles in a total of $2\times10^6$ cycles to a final value of $P^*=10^7$. The initial pressure of $P^*=1$ is small enough to allow for the fast de-correlation of the system from its starting configuration, while the step-wise exponential increase in pressure is a computationally efficient way of isolating the densest packing in the fewest number of MC cycles. As it has been previously shown for tetrahedra~\cite{Chen2010} and truncated tetrahedra~\cite{PabloACSNanot2012}, the densest packing obtained from this approach is independent of the starting configuration if a sufficient number of repeat simulations (e.g. with different seeds) are performed.  Here, we perform $100$ independent simulations for every truncation and system size, and record the densest packing within that $100$-run sample. The maximum packing fractions obtained from these 100 simulations are always within less than 0.5 per cent of the highest packing densities obtained from analytical and/or numerical solution of the packing problem, which suggests that the amount of sampling has been sufficient.

In order to assess the self-assembly potential of different tTBPs, MC simulations are performed in the canonical ensemble for packing fractions ranging from $55\%$ to $69\%$. We start each simulation from a configuration of $N=2\,048$ tTBPS rapidly compressed to a packing fraction of $\approx70\%$ and rescaled to the appropriate packing fractions. These simulations are performed for a minimum of $5\times10^7$ MC cycles. 
 
\subsection{The packing problem\label{subsec:packing}}
\label{subsection:packing_problem}
A collection of shapes $\{\mathcal{R}_i\}_{i=1}^{\infty}\subseteq\mathbb{R}^d$ is called a packing if their interiors are disjoint~\cite{BezdekConvexPacking2010}. For every packing, the packing fraction (or packing density) is defined as
\begin{eqnarray}
	\phi &:=& \limsup_{r\rightarrow\infty}\frac1{\text{Vol}[B(r)]}\sum_{i=1}^{\infty}\text{Vol}[\mathcal{R}_i\cap B(r)]\label{eq:back:packing},
\end{eqnarray}
with $B(r)=\{x\in\mathbb{R}^d:||x||\le r\}$. For a periodic packing of congruent objects, Eq.~(\ref{eq:back:packing}) takes the form
\begin{eqnarray}
\phi &=& \frac{NV_\mathcal{R}}{V_B},
\end{eqnarray}
where $N$ is the number of particles in the unit cell. $V_{\mathcal{R}}$ and $V_B$ are the volumes of the individual particle and the unit cell, respectively.

The packing problem searches for the densest possible packing and can be stated as follows: What is the arrangement of the shapes $\{\mathcal{R}_i\}_i$ that maximizes $\phi$ as defined in Eq.~(\ref{eq:back:packing})? In its most general formulation, without a confining box, the packing problem is an optimization problem with an infinite number of variables, and thus difficult to solve. However, since packings in spatially restricted systems depend only on a finite number of parameters, the local maxima of Eq.~(\ref{eq:back:packing}) can often be determined numerically or (in a few cases) even analytically. For instance, the densest lattice packings of three-dimensional compact convex shapes have been obtained using the classical method of Minkowski~\cite{Minkowski1904} for regular tetrahedra~\cite{Hoylman1970} and regular octahedra~\cite{Minkowski1904}. Betke and Henk developed an efficient computer algorithm for the determination of dense lattice packings of 3-polytopes, and applied it to Platonic and Archimedean solids~\cite{BetkeHenk2000}. Local maxima have been obtained for periodic non-lattice packings with a few particles in the unit cell for several three-dimensional objects~\cite{BezdekConvexPacking2010, ChenDSC2008, Kallus2010, TorquatoSuperballPRE2009, Chen2010}.

In this work, we use the following procedure proposed by Chen~\emph{et~al.}~\cite{Chen2010}.
Given a dense packing obtained from MC simulation, we identify all contacts between the particles in the unit cell. These contacts are then expressed as intersection equations, which act as constraints for the suboptimal optimization problem:
\begin{eqnarray}
	\label{eq:packing_opt}
	\begin{array}{lll}
		\text{minimize} & V(q_1,q_2,\cdots,q_n) \\
		\text{subject to} & I_k(q_1,q_2,\cdots,q_n)=0,~k=1,2,\cdots,m.
	\end{array}
\end{eqnarray}
where $V(q_1,q_2,\cdots,q_n)$ is the volume of the periodic box. The $q_i$ correspond to the $n$ degrees of freedom of the packing (examples are lattice vectors, positions and orientations of particles in the unit cell) and $I_k=0$ are the $m$ intersection constraints that need to be satisfied by the packing. Eq.~(\ref{eq:packing_opt}) is solved analytically or numerically to obtain the densest packing consistent with the constraints.

%%%%%%%%%%%%%%%%%%%%%%%%%%%%%%%%%%%%%%%%%%
%%%%%%%%%%%%%%%%%%%%%%%%%%%%%%%%%%%%%%%%%%
%%%%%%%%%%%                        R E S U L T S           %%%%%%%%%%%%%
%%%%%%%%%%%%%%%%%%%%%%%%%%%%%%%%%%%%%%%%%%
%%%%%%%%%%%%%%%%%%%%%%%%%%%%%%%%%%%%%%%%%%

\section{Results}
\subsection{Dense packings\label{subsec:densePackings}}
\begin{figure}
	\begin{center}
		\includegraphics[width=0.9\columnwidth]{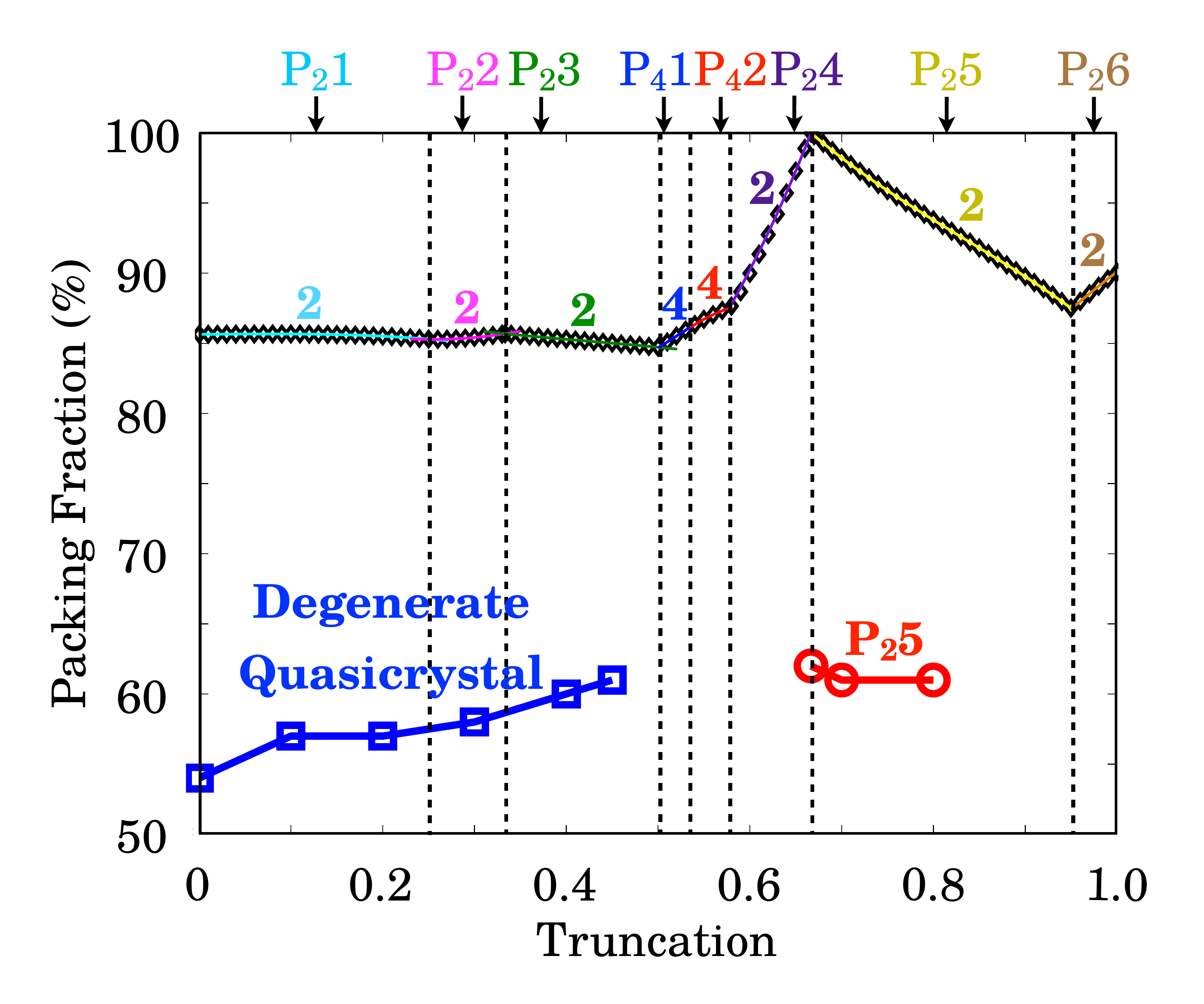}
		\caption{\label{fig:TBP:phimaxtTBP} (color online).~Packing fraction $\phi_{\max}(t)$ of the densest packings and self-assembled arrangements of tTBPs. The upper symbols are the maximum packing densities obtained from MC simulations, and the curves are calculated from the analytical expressions given in Appendix~\ref{appendix:tTBPanalytical} or numerical solutions outlined in Appendix~\ref{appendix:4pPackings}. In addition, the lower symbols correspond to the lowest densities at which each phase forms in self-assembly simulations.\label{fig:summary}
		}
	\end{center}
\end{figure}

The packing fractions determined from simulations with small boxes are reported in Fig.~\ref{fig:TBP:phimaxtTBP}. As with any continuous deformation, $\phi_{\max}(t)$ changes continuously with the truncation $t$. We identify eight distinct packing types by noting six discontinuities in the first derivative $\phi'_{\max}(t)$ and one discontinuity in the second derivative $\phi''_{\max}(t)$.
All packings have densities that exceed $84\%$, which shows that tTBPs are generally efficient packers. The tTBP with truncation $t=\frac23$ is a space-filling polyhedron previously indexed as $14$-III by Goldberg~\cite{Goldberg1979}. Eq.~(\ref{eq:packing_opt}) was formulated for all the packings. Details can be found in the Appendices~\ref{appendix:tTBPanalytical} and~\ref{appendix:4pPackings}. We label packings as P$_{n}i$, where $n$ specifies the number of particles in the unit cell and $i$ is a running index characteristic of the packing type. The packings and their interaction constraints are listed in Table~\ref{table:intersections}.

\begin{table}
	\caption{\label{table:intersections}The eight types of densest tTBP packings.  We list the numbers of independent intersection constraints for each packing.}
	\begin{tabular}{l|cccccccc}
		\hline\hline
		Type &~~P$_{2}1$~&~P$_{2}2$~&~P$_{2}3$~&~P$_{4}1$~&~P$_{4}2$~&~P$_{2}4$~&~P$_{2}5$~&~P$_{2}6$~\\
		\hline
		Vertex-face & 1 & 0 & 0 & 3 & 5 & 0 & 0 & 0\\
		Edge-edge & 2 & 2 & 2 & 2 & 2 & 4 & 2 & 3\\
		Edge-face & 0 & 0 & 0 & 2 & 2 & 0 & 0 & 0\\
		Face-face & 8 & 8 & 8 & 3 & 3 & 5 & 8 & 6\\
		\hline
		Total & 11 & 10 & 10 & 10 & 12 & 9 & 10 & 9\\
		\hline\hline
	\end{tabular}
\end{table}

\subsubsection{Two-particle packings\label{subset:twoppackings}}
Six of the eight densest packings have two particles in the minimal unit cell. Eq.~(\ref{eq:packing_opt}) was solved analytically for all six packings and the full mathematical description of each packing is given in Appendix~\ref{appendix:tTBPanalytical}. In each packing, the orientations of the two tTBPs are related by inversion symmetry, which means they are Kuperberg pairs~\cite{KuperbergDCG1990}. The layers formed by each particle will be denoted by + and -- for the original orientation and the inversion, respectively.

\begin{figure}
	\begin{center}
		\includegraphics[width=\columnwidth]{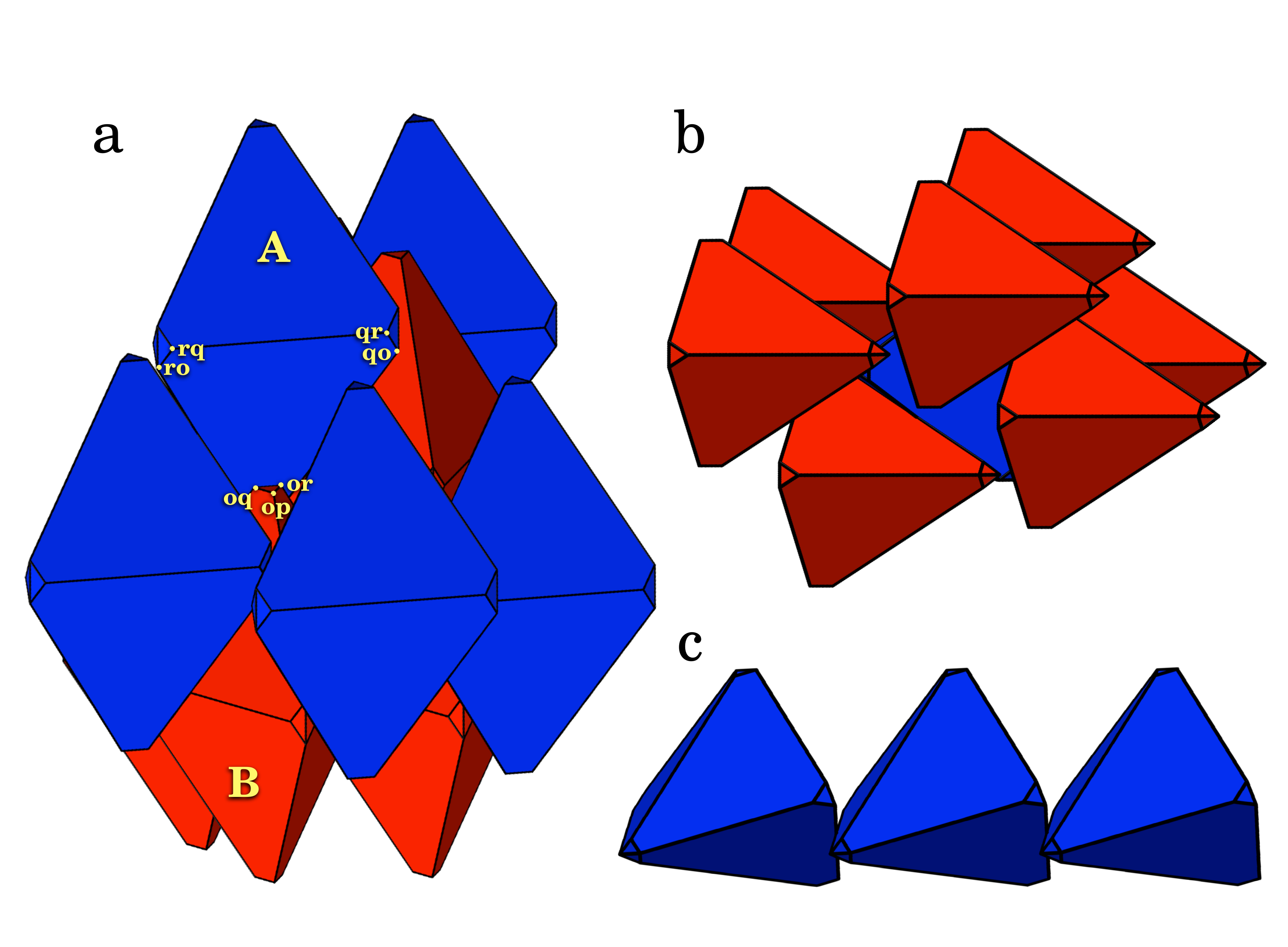}
		\caption{\label{fig:P2-1}
		(color online).~The packing P$_{2}1$ for the truncation $t=\frac15$. (a)~The polar triangle $F[\textbf{op},\textbf{oq},\textbf{or}]$ of particle B touches the peripheral hexagon $F[\textbf{o},\textbf{p},\textbf{q}]$ of particle A through the edge $E[\textbf{oq},\textbf{or}]$. (b)~The central blue particle has face-to-face contacts with eight of its neighbors, all of which are in an opposite layer. (c) Particles with identical orientations touch through edge-to-edge contacts only.
		}
	\end{center}
\end{figure}

For small truncations, we find a packing that is similar to the TBP crystal proposed in~\cite{Chen2010} except for a gradual shear to optimize the packing of slightly truncated TBPs. This packing type, which we denote by P$_{2}1$ (Fig.~\ref{fig:P2-1}a), has eleven independent intersection constraints that are given in Eq.~(\ref{eq:P$_{2}1$:intersections}), two more than the original TBP crystal~\cite{Chen2010}. The face-to-face contacts in the P$_{2}1$ packing are all among peripheral hexagons and neither the polar nor the equatorial triangles touch any other faces. All face-to-face contacts are between particles in opposite layers (Fig.~\ref{fig:P2-1}b) and particles with identical orientations touch through edge-to-edge contacts only (Fig.~\ref{fig:P2-1}c). The list of face-to-face contacts in P$_{2}1$ is identical to that in the TBP crystal. The maximum packing fraction for P$_{2}1$ is given by
\begin{eqnarray}
\phi_{2,1}(t) &=& \frac{500(8-3t^3)}{4671-30t-25t^2}.\label{eq:max_pf_P$_{2}1$}
\end{eqnarray}

\begin{figure}
	\begin{center}
		\includegraphics[width=\columnwidth]{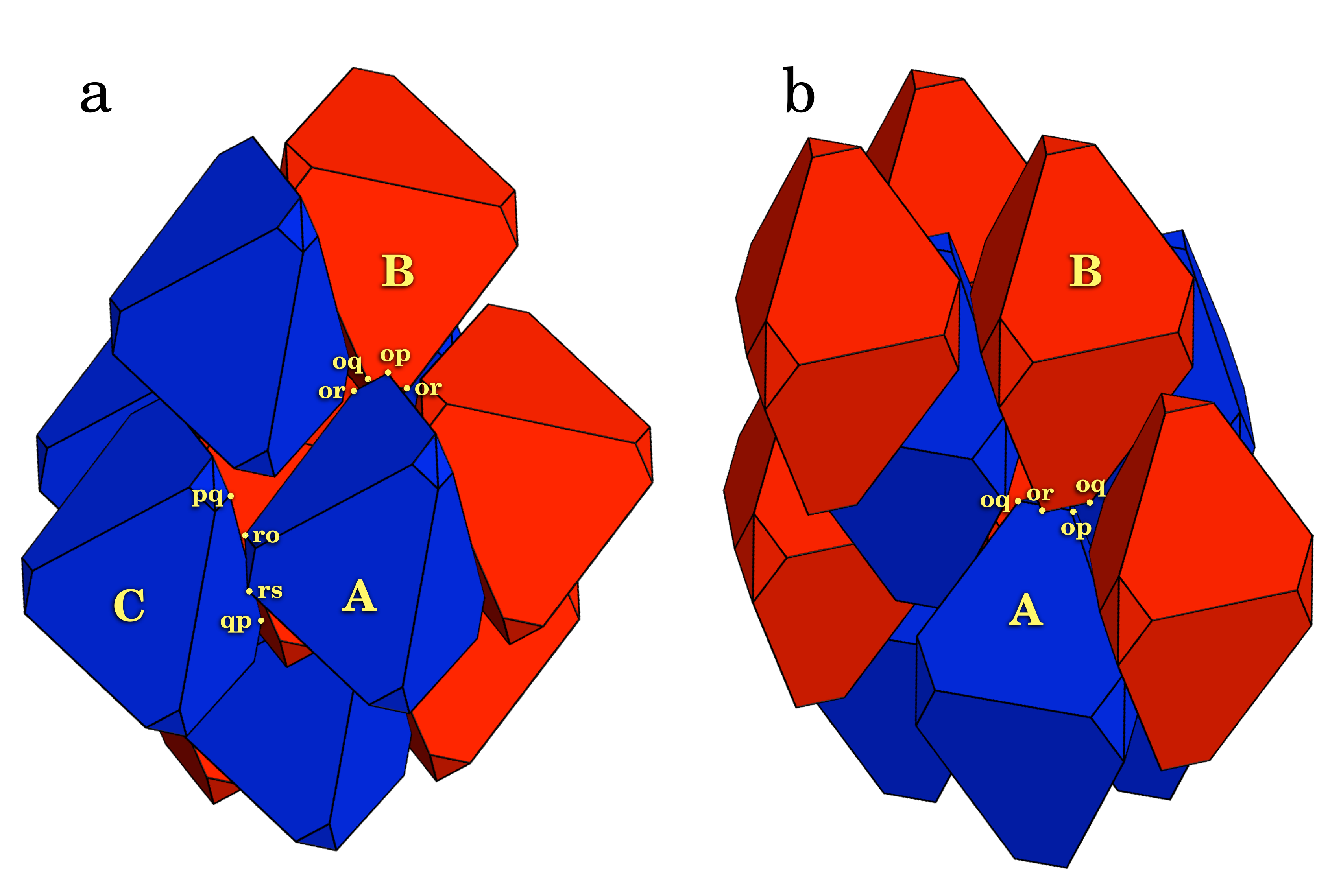}
		\caption{\label{fig:compare_P2_2_P2_3}
			(color online).~The packing P$_{2}2$ for the truncation $t=\frac3{10}$~(a) and the packing P$_{2}3$ for the truncation $t=\frac25$~(b). (a) In P$_{2}2$, the faces $F[\textbf{op},\textbf{oq},\textbf{or}]$ of the particles A and B touch and form a star-like pattern. The axial edge $E[\textbf{ro},\textbf{rs}]$ of particle A and the peripheral edge $E[\textbf{pq},\textbf{qp}]$ of particle C touch. (b) In P$_{2}3$, polar faces of particles A and B touch forming a star-like pattern. No axial edge of P$_{2}3$ touches an edge of a neighboring particle.
		}
	\end{center}
\end{figure}

The next two packing types, P$_{2}2$ (Fig.~\ref{fig:compare_P2_2_P2_3}a) and P$_{2}3$ (Fig.~\ref{fig:compare_P2_2_P2_3}b), are characterized by star-like patterns formed by the face-to-face touching of polar triangles of particles in opposite layers (Fig.~\ref{fig:compare_P2_2_P2_3}). This is in contrast to P$_{2}1$ where the polar triangles only touch the peripheral hexagons through one of their edges (Fig.~\ref{fig:P2-1}a). Both P$_{2}2$ and P$_{2}3$ have a total of ten independent intersection constraints given in Eq.~(\ref{eq:P$_{2}2$:intersection}) and Eq.~(\ref{eq:P$_{2}3$:intersection}), respectively. The major difference between these two otherwise similar packings is the existence of an edge-to-edge contact in P$_{2}2$, where an axial edge (an edge parallel to the three-fold axis of the particle) touches a peripheral edge (Fig.~\ref{fig:compare_P2_2_P2_3}a). The maximum packing fractions of P$_{2}2$ and P$_{2}3$ are given by
\begin{eqnarray}
	\phi_{2,2}(t) &=& \frac{27(8-3t^3)}{283+131t-319t^2+64t^3} \label{eq:max_pf_P$_{2}2$},\\
	\phi_{2,3}(t) &=& \frac{27(8-3t^3)}{2(1+t)(119-80t+8t^2)}. \label{eq:max_pf_P$_{2}3$}
\end{eqnarray}

\begin{figure}
	\begin{center}
		\includegraphics[width=\columnwidth]{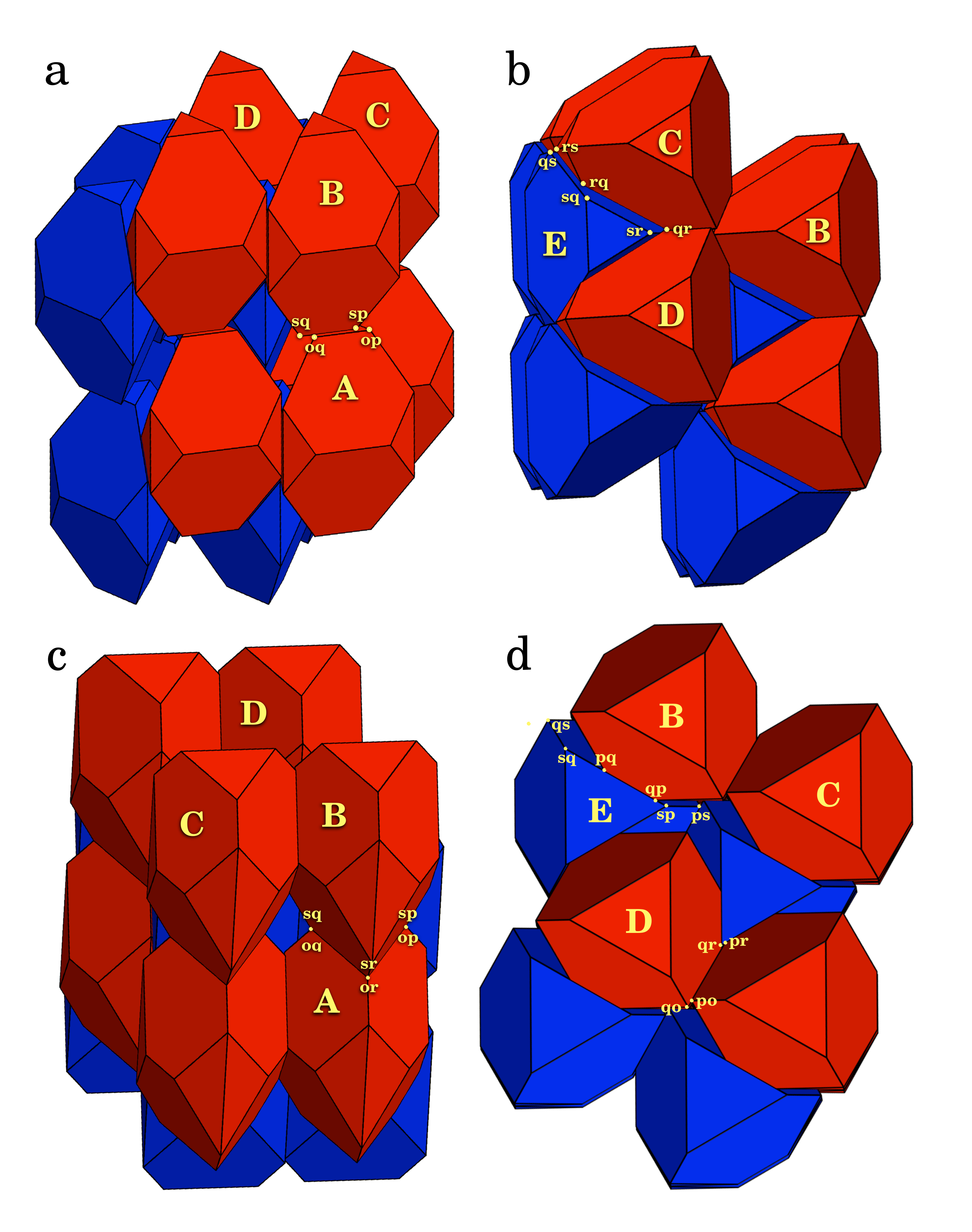}
		\caption{\label{fig:compare_P2_4_P2_5}
		(color online).~The packing P$_{2}4$ for the truncation $t=\frac35$~(a,b) and the packing P$_{2}5$ for the truncation $t=\frac45$~(c,d). (a,c)~The side view reveals that the face $F[\textbf{op},\textbf{oq},\textbf{or}]$ of particle A touches the face $F[\textbf{sp},\textbf{sq},\textbf{sr}]$ of particle B. The contact is shifted in~(a) but not in~(c). Equatorial triangles of particles B, C and D come together in~(b,d). It is visible in the top view~(d) that particle D shifts away from the (B,C,D) triplet as $t$ becomes larger.}
	\end{center}
\end{figure}

The next two types of two-particle packings, P$_{2}4$ (Fig.~\ref{fig:compare_P2_4_P2_5}a-b) and P$_{2}5$ (Fig.~\ref{fig:compare_P2_4_P2_5}c-d), are characterized by the contact of polar triangles of identically oriented particles (Fig.~\ref{fig:compare_P2_4_P2_5}a,c).  This is unlike P$_{2}2$ and P$_{2}3$ where star-like patterns form between polar triangle of particles in alternate layers (Fig.~\ref{fig:compare_P2_2_P2_3}). In both P$_{2}4$ and P$_{2}5$, particles arrange into a hexagonal lattice in each layer with their equatorial triangles touching. Face-to-face contacts between particles in opposite layers occur through peripheral hexagons. Therefore, the + and -- layers are shifted with respect to one another and the entire structure is isostructural to a stretched hexagonally close-packed (hcp) lattice (Fig.~\ref{fig:compare_P2_4_P2_5}b,d). 

There are two major structural differences between these two packings. The face-to-face contacts between polar triangles are perfect in P$_{2}5$ (Fig.~\ref{fig:compare_P2_4_P2_5}c), while the touching polar triangles are shifted in P$_{2}4$ (Fig.~\ref{fig:compare_P2_4_P2_5}a). The contacts between equatorial triangles are also different. In P$_{2}4$, equatorial triangles of triplets of particles come together, like for example B, C and D in Fig.~\ref{fig:compare_P2_4_P2_5}b.  Yet, in P$_{2}5$ one of each of the three particles in those triplets start distancing from the other two as the truncation parameter increases. This leads to an elongation of the intra-layer hexagonal lattice along one of its lattice vectors (Fig.~\ref{fig:compare_P2_4_P2_5}d). Overall, P$_{2}4$ and P$_{2}5$ have nine and ten independent intersection constraints that are given by Eq.~(\ref{eq:P$_{2}4$:intersection}) and Eq.~(\ref{eq:P$_{2}5$:intersection}) respectively. The maximum packing fractions are given by
\begin{eqnarray}
	\phi_{2,4}(t) &=& \frac{2(8-3t^3)}{3(2-t)(4+4t-7t^2)} \label{eq:max_pf_P$_{2}4$},\\
	\phi_{2,5}(t) &=& \frac{28(8-3t^3)}{3(2-t)(-4+124t-65t^2)}. \label{eq:max_pf_P$_{2}5$}
\end{eqnarray}

The boundary between the P$_{2}4$ region and the P$_{2}5$ region at $t=\frac23$ corresponds to a tTBP that tiles Euclidean space (Fig.~\ref{fig:spf_figs}a). This space-filling polyhedron is known and was indexed as $14$-III by Goldberg in 1979~\cite{Goldberg1979}. We rediscover this polyhedron in our study and observe that the corresponding tiling forms in self-assembly simulations of the same building block (see below in Fig.~\ref{fig:spf_figs}b-c). 

\begin{figure}
	\begin{center}
		\includegraphics[width=\columnwidth]{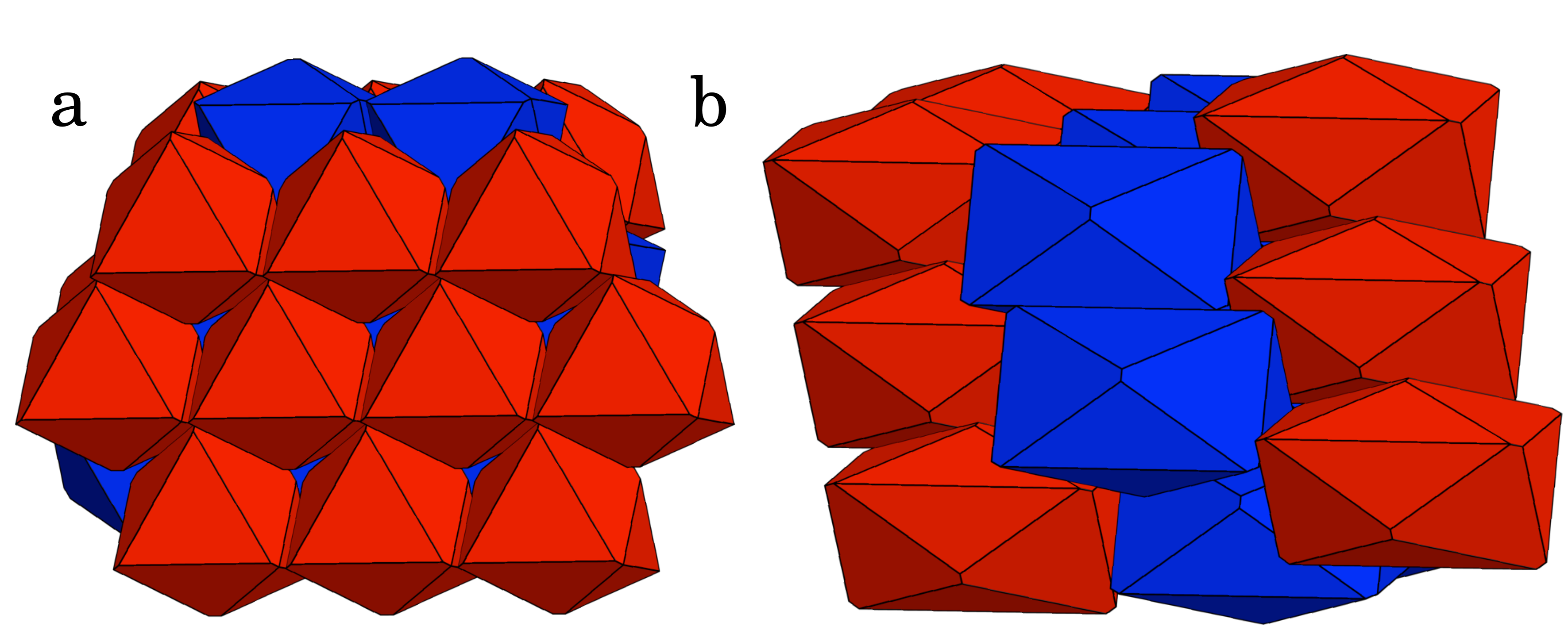}
		\caption{\label{fig:P2-6}
		(color online).~(a) Top and (b) side view of the packing P$_{2}6$ for the truncation $t=\frac{24}{25}$.
		}
	\end{center}
\end{figure}

The last type of two-particle packings, P$_{2}6$, is characterized by the absence of contacts between two consecutive + and -- layers. Like P$_{2}2$ and P$_{2}3$, polar triangles of particles in opposite layers intersect, but the contact is less perfect and the particles tend to have a relatively large lateral shift (Fig.~\ref{fig:P2-6}b). This leads to a staircase arrangement for each layer. Like P$_{2}4$, the equatorial triangles of particles in identical layers are face-to-face (Fig.~\ref{fig:P2-6}a). The peripheral hexagons also touch the corresponding hexagons from opposite layers.  This packing has a total of nine independent intersection constraints given by Eq.~(\ref{eq:P$_{2}6$:intersection}). Its packing fraction is given by
\begin{eqnarray}
	\phi_{2,6}(t) &=& \frac{20(8-3t^3)}{3(2-t)(52+20t-35t^2)}. \label{eq:max_pf_P$_{2}6$}
\end{eqnarray}

Further information about these packings, including analytical expressions for lattice vectors and intersection constraints, are given in Appendix~\ref{appendix:tTBPanalytical}.

\subsubsection{Four-particle packings}
\begin{figure}
	\begin{center}
		\includegraphics[width=\columnwidth]{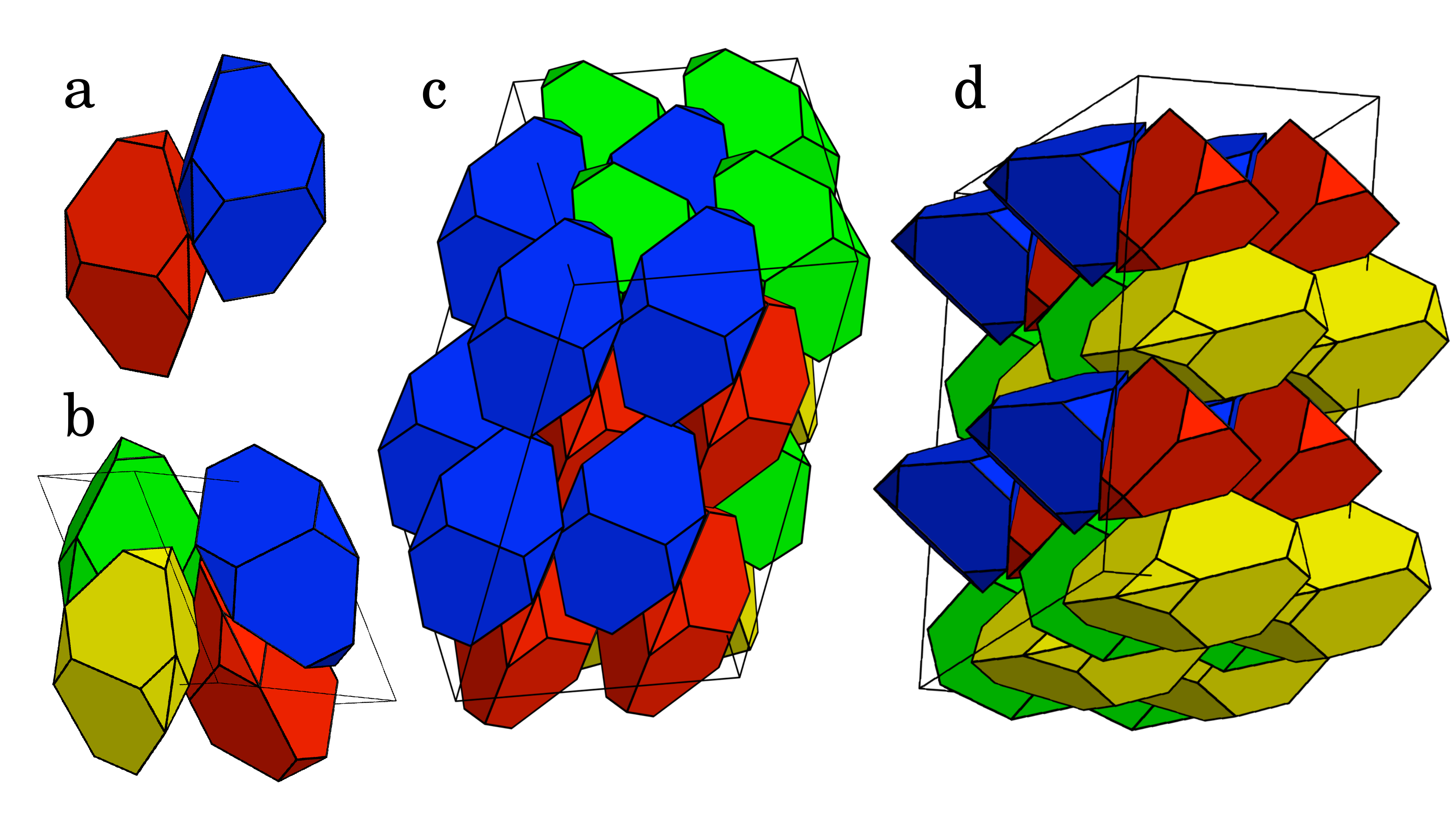}
		\caption{\label{fig:TBP:4ppacking}
			(color online).~Dense packings of tTBPs with four particles in the unit cell for the truncation $t=\tfrac{11}{20}$. (a)~The dimer is a Kuperberg pair and the basic building blocks of the packings P$_{4}1$ and P$_{4}2$. The red and blue tTBPs belong to the same dimer, while the green and yellow particles belong to the other dimer rotated by an angle of about $60^{\circ}$. The crystal structure is depicted from different perspectives in (b-d). 
		}
	\end{center}
\end{figure}

\begin{figure}
	\begin{center}
		\includegraphics[width=0.9\columnwidth]{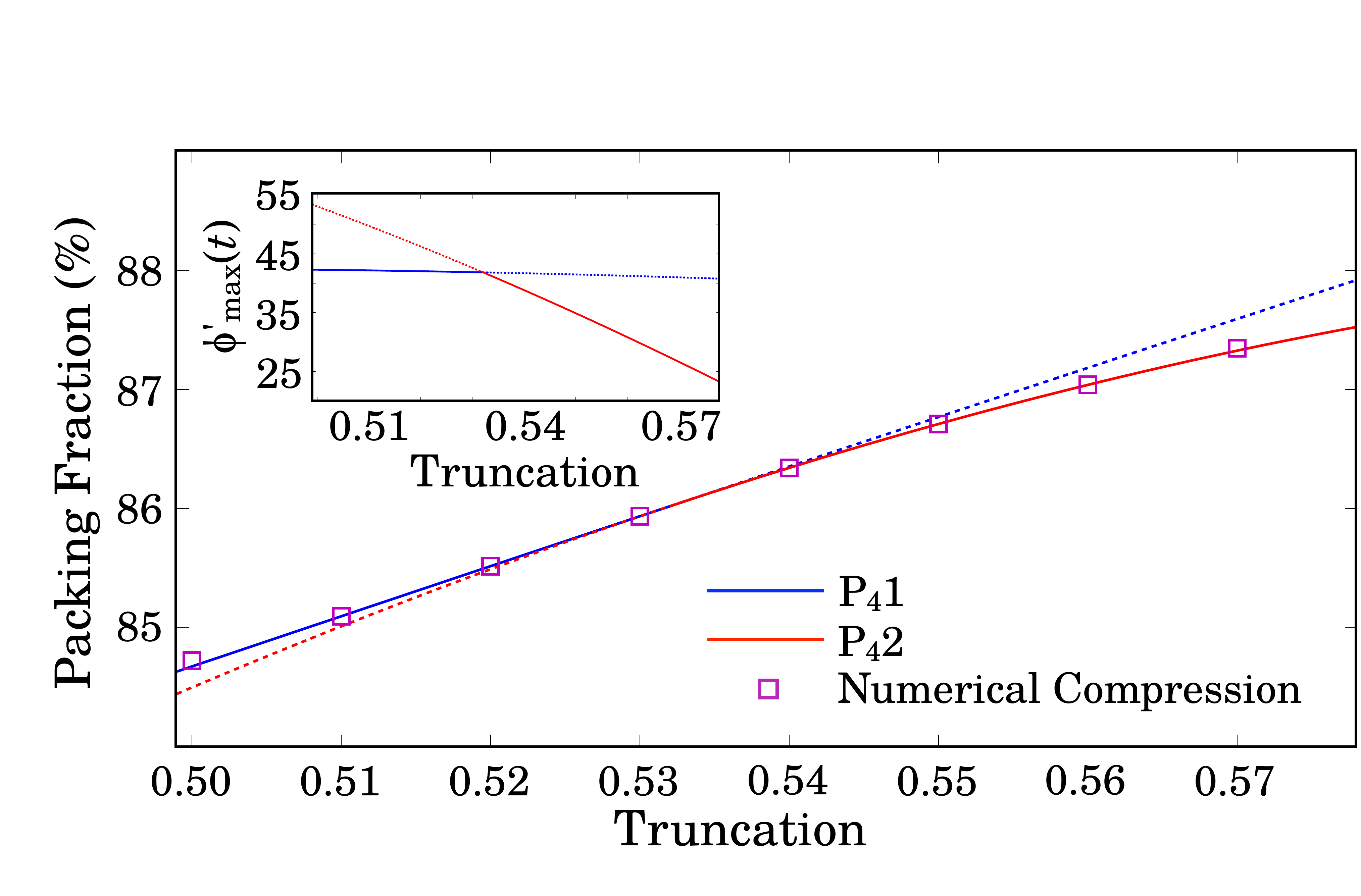}
		\caption{\label{fig:4p_tr}(color online).~The packing fraction $\phi_{\max}(t)$ and its derivative $\phi'_{\max}(t)$ are continuous at the transition from P$_{4}1$ to P$_{4}2$. Magenta squares are the densest packings obtained from small-box MC simulations. The blue and the red curves are the packing fractions calculated from numerical solution of Eq.~(\ref{eq:packing_opt}). These curves are depicted in solid lines wherever they correspond to the densest packing and dashed otherwise.  As can be seen in the inset, there is a discontinuity in the slope of $\phi'_{\max}(t)$ at $t_4\approx0.5321$, which corresponds to a second-order transition in the truncation space.}
	\end{center}
\end{figure}

The two packing types observed for the intermediate truncations $t_3\approx0.5010\le t\le t_5\approx 0.5776$ have four particles in the fundamental domain. The basic building block of these packings is a (non-convex) dimer of tTBPs with partially touching peripheral hexagons (Fig.~\ref{fig:TBP:4ppacking}a). Each dimer is a Kuperberg pair as the orientations of its constituent tTBPs are related by inversion. The fundamental domain contains two such dimers that are rotated with respect to one another at an angle of about $60$ degrees. All face-to-face contacts are between particles in Kuperberg pairs with identical orientations: Polar triangles of particles with identical orientations touch (Fig.~\ref{fig:TBP:4ppacking}c), while four out of the six peripheral hexagons of each particle touches the peripheral hexagons of neighbors with opposite orientations, \emph{i.e.}\ the neighbors with orientations related to the central particle by inversion (Fig.~\ref{fig:TBP:4ppacking}d). In contrast, the contacts between particles in separate dimer types are either edge-to-edge or vertex-to-face (Fig.~\ref{fig:TBP:4ppacking}b). There are a total of ten and twelve independent intersection constraints for P$_{4}1$ and P$_{4}2$, respectively, alongside eight additional symmetry constraints given by Eq.~(\ref{eq:appendix:symm_const}).  Due to the non-linear nature of the intersection constraints, no analytical solution can be obtained for Eq.~(\ref{eq:packing_opt}). Numerical solutions were however, obtained and the results are depicted in Figs.~\ref{fig:summary} and~\ref{fig:4p_tr}.

These two packings are similar to the SM2 packing of hard ellipsoids~\cite{DonevStillingerPRL2004,schillingPRE2007}, which has a monoclinic unit cell with two ellipsoids forming an angle between them. The SM2 phase is denser than the stretched face-centered cubic (fcc) phase, which is obtained from an affine transformation of the sphere fcc packing~\cite{MulderFrenkelMolPhys1985}. For sufficiently large aspect ratios, the SM2 packing is always more stable than the stretched fcc phase~\cite{RaduSchilling2009}. 

A remarkable observation about the two four-particle packings is that all intersection constraints satisfied by P$_{4}1$ are also satisfied by P$_{4}2$. In addition, P$_{4}2$ satisfies two additional constraints that correspond to two inter-dimer vertex-to-face contacts. This makes the set of all feasible packings for P$_{4}2$ a \emph{subset} of all feasible packings for P$_{4}1$, and leads to a unique transition of the packing behavior in the truncation space. While both $\phi_{\max}(t)$ and $\phi'_{\max}(t)$ change continuously from P$_{4}1$ to P$_{4}2$, a discontinuity occurs only in the second derivative (Fig.~\ref{fig:4p_tr}). We therefore characterize this transition as a `second-order' transition in the truncation space. Analogously, the usual behavior of a discontinuity in $\phi'_{\max}(t)$ would be called a `first-order' transition.

\subsection{Self-assembly simulations\label{subsec:self_assembly}}
The tTBP family assembles far fewer ordered structures than the truncated tetrahedron family studied in Ref.~\cite{PabloACSNanot2012}. We observe the assembly of only two distinct crystal structures. A dodecagonal quasicrystal forms for truncations $t\le 0.45$. We already reported the formation of this quasicrystal in a system of non-truncated TBPs~\cite{HajiAkbariDQC2011}. At higher truncation, P$_{2}5$ forms in the range $\frac23=0.66\cdots\le t\le0.80$. For the remaining truncations, no spontaneous ordering was observed on the (long) timescale of our simulations.

We analyze the structures obtained from our simulations using the tools outlined in Ref.~\cite{HajiAkbariDQC2011}.  Table.~\ref{table:TBP:tTBP_DQC} lists the minimum packing fraction at which the degenerate quasicrystal forms for each truncation as well as the number of MC cycles it took for it to form in our simulations. As expected and as also observed with imperfect tetrahedra~\cite{DamascenoImperfect}, it becomes increasingly difficult for tTBPs to form the quasicrystal as the truncation increases and higher packing fractions and longer simulation times are necessary for crystallization.
\begin{table}
	\begin{center}
		\caption{\label{table:TBP:tTBP_DQC} Formation of the degenerate quasicrystal of tTBPs. $\phi_{\min}$ corresponds to the lowest packing fraction at which the quasicrystal assembles from the disordered fluid. $\phi_{\max}$ is to the maximum packing fraction obtained from a numerical compression of the assembled bulk quasicrystal.}
		\begin{tabular}{cccc}
			\hline\hline
			Truncation &~~~~~$\phi_{\min}~~~~~$ &~~~~~MC Cycles~~~~~&~~~~~$\phi_{\max}$~~~~~\\
			\hline
			$0.1$ & $57\%$ & $4\times10^7$ & $81.21\%$\\
			$0.2$ & $57\%$ & $6\times10^7$ & $80.93\%$\\
			$0.3$ & $58\%$ & $9\times10^7$ & $79.74\%$\\
			$0.4$ & $60\%$ & $15\times10^7$ & $81.31\%$\\
			$0.45$ & $61\%$ & $28\times10^7$ & $80.06\%$ \\
			\hline
		\end{tabular}
	\end{center}
\end{table}

\begin{figure*}
	\begin{center}
		\includegraphics[width=1.8\columnwidth]{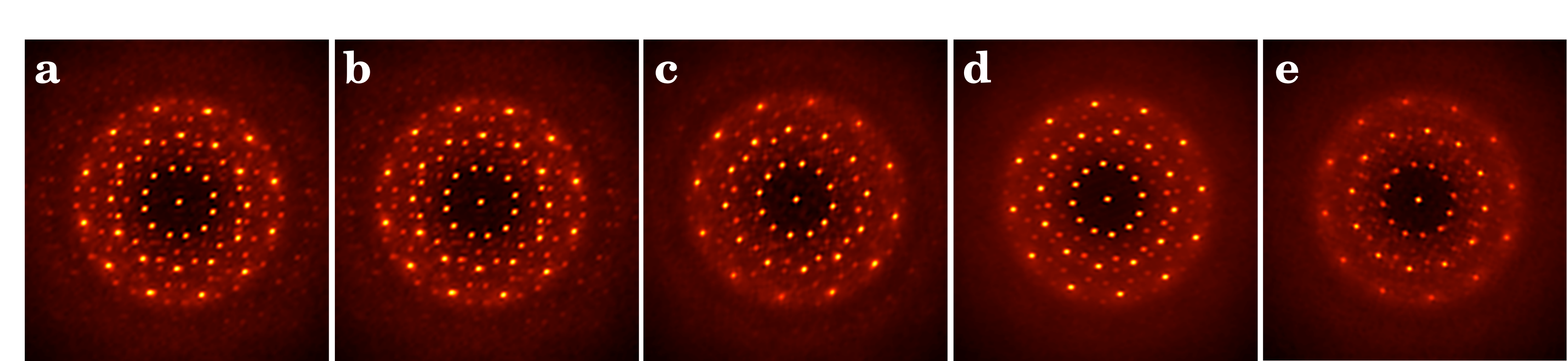}
		\caption{\label{fig:TBP:tTBP_diff}
		(color online).~Diffraction images of the degenerate dodecagonal quasicrystals of tTBPs for the truncations (a) $t=0.1$, (b) $t=0.2$, (c) $t=0.3$, (d) $t=0.4$, and (e) $t=0.45$. The sharpness and intensity of the peaks decreases slowly with truncation.
		}
	\end{center}
\end{figure*}

To compare the structural quality of the quasicrystals to the degenerate quasicrystal obtained for zero truncation, we calculate the diffraction patterns for the assembled structures compressed to their maximum packing fractions (so-called "inherent structures"~\cite{StillingerWeberJCP1985}). Due to the degenerate nature of the quasicrystal, the scatterers are placed not at the centroids of TBPs, but instead at the centroids of each of the two tetrahedra that comprise each non-truncated TBP. The resulting diffraction patterns resemble the ones obtained for a system of hard tetrahedra~\cite{HajiAkbariEtAl2009}. The twelvefold symmetry in the diffraction pattern, which is the signature property of a dodecagonal quasicrystal, is readily visible in all diffraction patterns (Fig.~\ref{fig:TBP:tTBP_diff}). It is notable that neither the quality of the diffraction patterns nor the maximum (post-compression) packing fractions of the quasicrystal(s) are significantly different from the degenerate quasicrystals formed by untruncated TBPs. This suggests that the structural quality of the quasicrystal is only slightly compromised as a result of truncating the building blocks, even though it gets gradually harder to assemble.

\begin{figure}
	\begin{center}
		\includegraphics[width=\columnwidth]{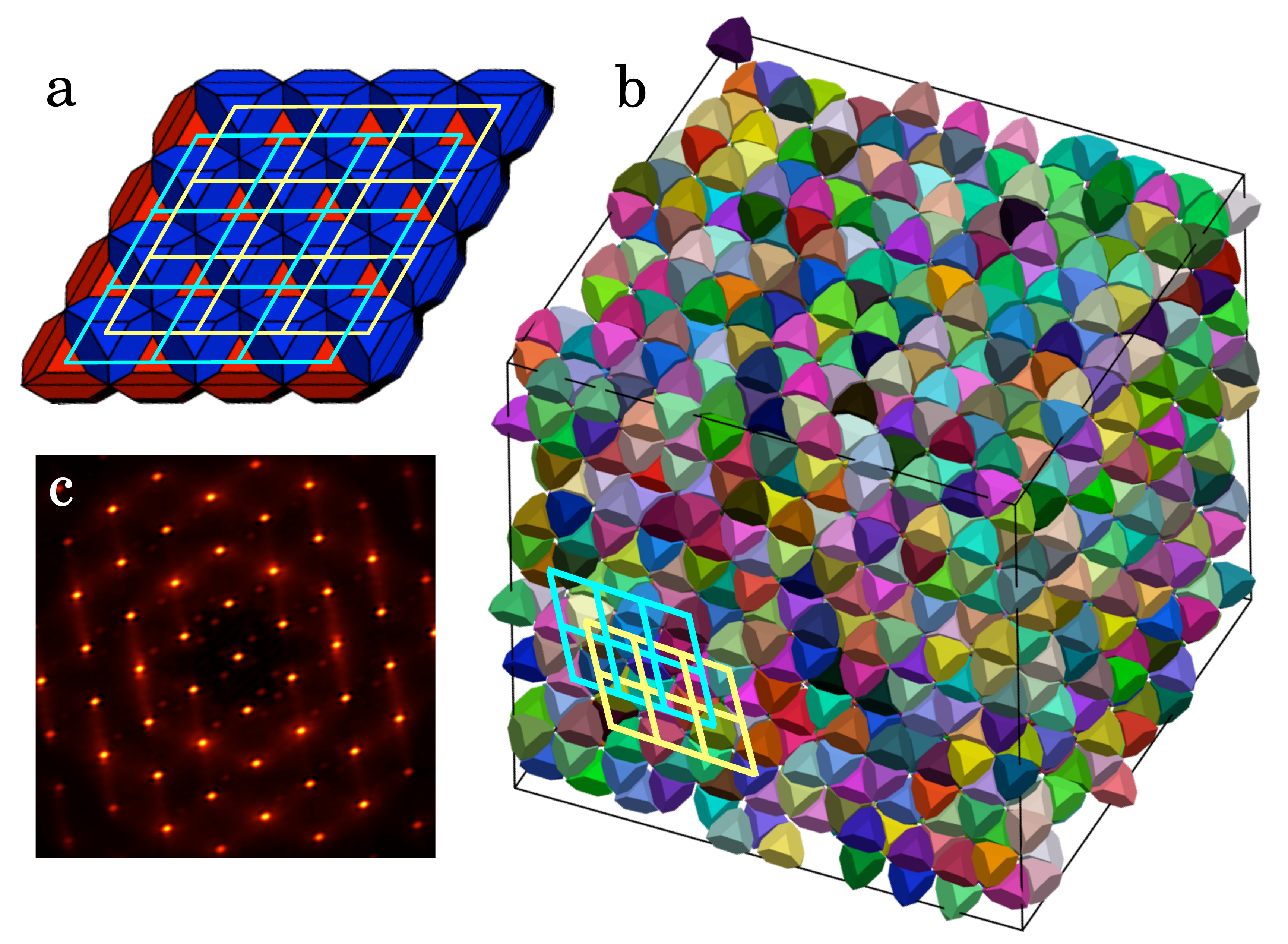}
		\caption{\label{fig:spf_figs} (color online).~(a) Geometrically constructed space-filling packing of the tTBP system for  $t=\frac23$. (b) The same structure is formed in a simulation of $2\,048$ tTBPs in an isochoric simulation at $\phi=62\%$. Colors are chosen randomly. (c) The diffraction pattern of (b) calculated from placing scatterers at the centers of tTBPs. Little peak broadening and diffuse scattering is present, which suggests the assembly is essentially perfect.}
	\end{center}
\end{figure}

For truncations $\frac23=0.66\cdots\le t\le0.80$, the system assembles into a simple crystalline structure identical to the P$_{2}5$ packing (described in section~\ref{subset:twoppackings}). Fig.~\ref{fig:spf_figs}b shows the final snapshot of an isochoric simulation of $2\,048$ tTBPs with $t=\frac23$ at $\phi=62\%$. In Fig.~\ref{fig:spf_figs}a, we show the analytically constructed P$_{2}5 $ packing for the same truncation, which is one of Goldberg's space-filling tilings~\cite{Goldberg1979}. In both panels, the location of the particles in the + and -- layers are highlighted with yellow and cyan lines, respectively (color online). Fig.~\ref{fig:spf_figs}c shows the diffraction pattern calculated when viewed along the vector perpendicular to the layers; the scatterers are positioned at the cetroids of tTBPs and not the constituent truncated tetrahedra. The observed sixfold symmetry is consistent with the elongated hcp structure of the P$_{2}5$ packing. For comparison, the truncated tetrahedron system at $t=\frac23$ assembles into a diamond crystal~\cite{PabloACSNanot2012}. The packing observed in this study is not degenerate to the diamond crystal observed for truncated tetrahedra. 

\subsection{The role of particle shape and local order\label{subsec:shapeLocalOrder}}

Except for P$_25$, none of the dense packings found in this study form in self-assembly simulations.  As mentioned in Section~\ref{section:intro}, the densest packing of every hard particle will be thermodynamically stable eventually for sufficiently large pressures. As previously shown for  tetrahedra~\cite{HajiAkbaricondmat2011}, truncated tetrahedra~\cite{PabloACSNanot2012}, triangular bipyramids~\cite{HajiAkbariDQC2011}, and several other shapes~\cite{PabloScience2012}, hard particles can assemble at intermediate densities into structures that are very different from their densest packings. While it has been observed that the assembled packings are typically similar to the dense disordered fluid on a local level~\cite{PabloScience2012}, no such relation is known between the fluid or the assembled packings and the densest packings.

There are several ways to quantify the similarity of two packings on a local level. The simplest approach is to use the coordination number (CN) by integrating the radial distribution function $g(r)$,
\begin{eqnarray}
\text{CN} &=& \int_{0}^{r_0}4\pi\rho_n r^2g(r)dr.
\end{eqnarray}
with a cut-off $r_0$ at the minimum between the first and second peak. Here $\rho_n$ is the average number density. It has been demonstrated recently by Damasceno~\emph{et~al.}~\cite{PabloScience2012} that there is a nearly perfect correlation between the CN of the dense disordered fluid and the CN of the self-assembled structure. We calculate CN for the dense disordered fluid, the densest packing, and the self-assembled structures at $\phi=60\%$. The results are presented in Table~\ref{table:Nc}. All assembled structures have CNs that are close to that of the disordered fluid at the same density. It is noteworthy that the proximity of CNs of the fluid and a particular packing is not a sufficient condition for self-assembly to occur since CN is only a spatially averaged measure of local order and does not take into account the rotational anisotropy of the nearest-neighbor shell. Moreover, CN can be problematic for certain elongated shapes, such as ellipsoids since it can underestimate the local coordination shell. However when there are stark differences in the CNs of the fluid and the densest packing, for example at $t\ge0.85$, the corresponding densest packing never forms in our self-assembly simulations.
\begin{table*}
	\caption{\label{table:Nc}Coordination number (CN) for the dense disordered fluid, the self-assembled structure, and the densest packing at different values of truncation. The isoperimetric quotient (IQ) of a particle $p$ is defined as $\text{IQ}=36\pi V_p^2/S_p^3$. It is a measure of the particle sphericity~\cite{PabloScience2012}.}
	\begin{tabular}{ccccccc}
		\hline\hline 
		~~Truncation~~~~ &~~~IQ~~~ &~~~~Densest Packing~~~~&~~~~Self-assembled Structure~~~~&~~~~CN$_{\text{fluid}}$~~~~&~~~~CN$_{\text{self-assembled}}$~~~~&~~~~CN$_{\text{packing}}$~~~\\
		\hline
		$0.10$ & 0.3623 & P$_{2}1$ & Quasicrystal &  1.9 & 1.9 & 1.7 \\
		$0.20$ & 0.3735 & P$_{2}1$ & Quasicrystal & 1.9 & 1.9 & 1.5  \\
		$0.30$ & 0.3913 & P$_{2}2$ & Quasicrystal & 2.0 & 1.9 & 4.0\\
		$0.40$ & 0.4150 & P$_{2}3$ & Quasicrystal & 2.0 & 1.9 & 2.0\\
		$0.45$ & 0.4290 & P$_{2}3$ & Quasicrystal & 2.0 & 1.9 & 1.9 \\
		$0.50$ & 0.4444 & P$_{2}3$ & -- & 2.0 & -- & 1.9 \\
		$0.55$ & 0.4611 & P$_{4}2$ & -- & 2.0 & -- & 1.6\\
		$0.60$ & 0.4791 & P$_{2}4$ & -- & 2.2 & -- & 3.9 \\
		$0.67$ & 0.5061 & P$_{2}5$ & P$_{2}5$ & 4.6 & 3.8 & 3.8 \\
		$0.70$ & 0.5182 & P$_{2}5$ & P$_{2}5$ & 4.5 & 3.9 & 3.9 \\
		$0.80$ & 0.5600 & P$_{2}5$ & P$_{2}5$ & 4.0 & 4.2 & 4.2 \\
		$0.85$ & 0.5808 & P$_{2}5$ & -- & 1.8 & -- & 5.1\\
		$0.90$ & 0.6004 & P$_{2}5$ & -- & 1.7 & -- & 2.7 \\
		$0.95$ & 0.6176& P$_{2}5$ & --& 1.5 & -- & 2.5 \\
		$1.00$ & 0.6304 & P$_{2}6$ & --& 1.4 & -- & 2.6 \\
		\hline
	\end{tabular}
\end{table*}

\begin{figure}
	\begin{center}
		\includegraphics[width=0.9\columnwidth]{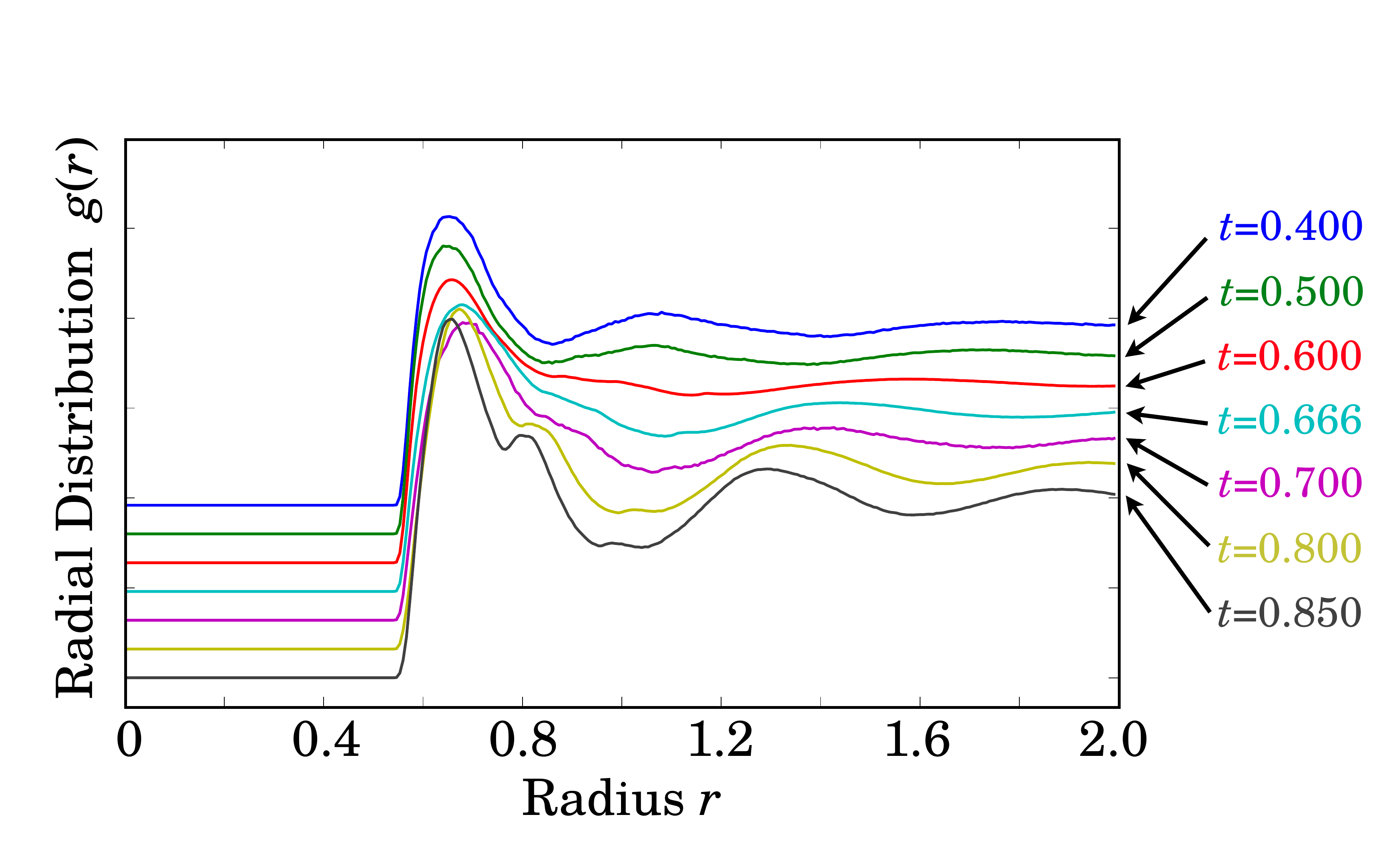}
		\caption{\label{fig:rdf_fluid}(color online).~Radial distribution functions of disordered fluids of tTBPs with various truncations at packing fraction $\phi=60\%$. The curves are shifted vertically for clarity.}
	\end{center}
\end{figure}

\begin{figure}
	\begin{center}
		\includegraphics[width=0.9\columnwidth]{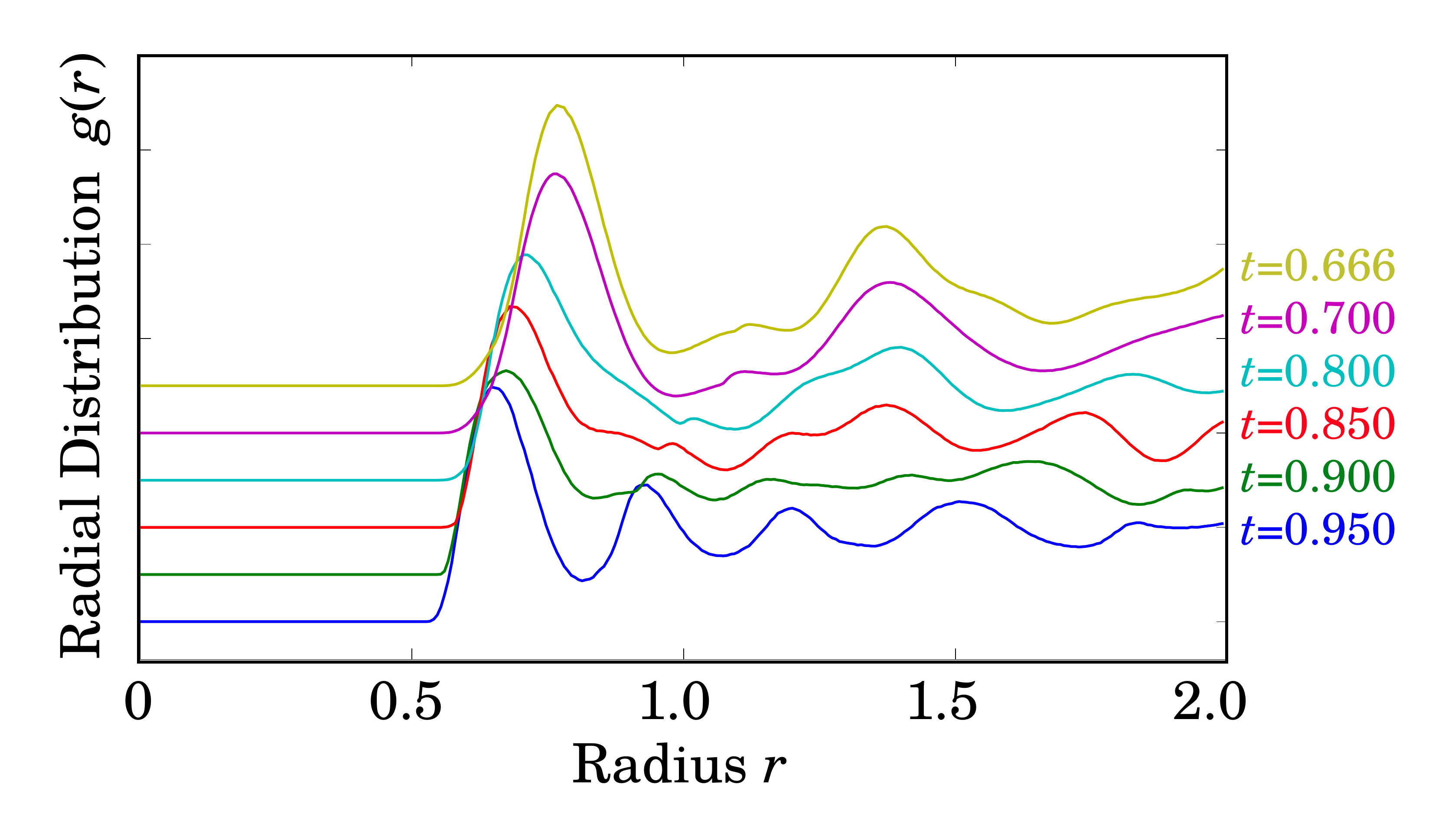}
		\caption{\label{fig:rdf_P2_5}
		(color online).~Radial distribution functions calculated from MC simulations of a system of $4\times4\times4$ unit cells of the P$_25$ packing at $\phi=60\%$.
		}
	\end{center}
\end{figure}

Further, we observe an abrupt change in the CN of the disordered fluid at around $t=0.6$. This change can be understood by directly inspecting the behavior of the radial distribution function of the disordered fluid. As can be seen in Fig.~\ref{fig:rdf_fluid}, the first peak of $g(r)$ grows and broadens for truncations from $t=0.66$ to $t=0.80$, which increases CN for these truncations. Near $t=0.85$ a small second peak appears at $r\approx 0.8$, which reduces CN by our definition and may explain why P$_25$ fails to form for truncations beyond $0.8$, despite being the densest packing for truncations as high as $0.95$. A similar peak-splitting occurs in P$_25$ near $t=0.90$ (Fig.~\ref{fig:rdf_P2_5}) which leads to an abrupt decrease in CN$_{\text{packing}}$ from $5.1$ to $2.7$. This is in line with our earlier observation that layers are stretched along one lattice direction as $t$ increases. The shifting of particles in the triplets of tTBPs sharing peripheral hexagons (explained in Section~\ref{subset:twoppackings}) makes the structure less uniform, which might also be a contributing factor in its kinetic inaccessibility. 

Damasceno~\emph{et~al.}~\cite{PabloScience2012} also find a correlation between the isoperimetric quotient~(IQ) of a crystal-forming building block and the type of crystal that it assembles. Crystal formers with intermediate IQs tend to form non-Bravais lattices while those with large IQs are more likely to form face-centered and body-centered cubic lattices. This is consistent with our findings in Table~\ref{table:Nc} as the particles forming the quasicrystal all have mostly intermediate isoperimetric quotients and the particles forming the P$_25$ packing have IQs compatible with simple crystal formation.

\section{Discussion and conclusions\label{sec:conclusion}}

We found that the formation of the degenerate dodecagonal quasicrystal from truncated triangular bipyramids is robust and can be observed for truncations as high as $0.45$. This is in line with earlier studies of the truncated tetrahedron system, and might be a consequence of the close local structural similarity of the quasicrystal and the disordered fluid. The quasicrystalline phase has been shown to be more favorable thermodynamically than the densest known packings for tetrahedra~\cite{HajiAkbaricondmat2011} and triangular bipyramids~\cite{HajiAkbariDQC2011} at intermediate densities.
Whether the same is true for tTBPs or truncated tetrahedra, is beyond the scope of this work and can be the subject of future studies.

The question of the kinetics of quasicrystal formation in these systems is undoubtedly interesting and open for future exploration. Our studies suggest that the self-assembly process is robust and is not affected significantly by small symmetric truncations of the vertices of the building blocks. This robustness is especially promising from an experimental viewpoint, because it is difficult to produce geometrically perfect shapes on nanometer to micron scales. Our finding hence increases the prospects of observing a colloidal quasicrystal for imperfectly shaped triangular bipyramids.

In all earlier explorations of densest packings obtained for a continuously transforming shape~\cite{TorquatoSuperballPRE2009, DijkstraBowlPRE2010, KallusPRE2011,PabloACSNanot2012,Dijkstra2012}, packing types adjacent in the density plot are conjoined via discontinuities in the first derivative of $\phi_{\max}$ with respect to the variable used for the continuous transformation of that shape. All two-particle packings in this study are indeed characterized by discontinuities in $\phi'_{\max}(t)$. In contrast, the transition from P$_41$ to P$_42$ is unique in the sense that a discontinuity occurs in $\phi''_{\max}(t)$, but not in $\phi'_{\max}(t)$. With breakthroughs in the synthesis of faceted nanoparticles that has spurred interest in studying packings for continuously transforming building blocks~\cite{GangPRL2011,Dijkstra2012}, it is likely that more second- and higher order transformations will be discovered for other families of continuously transforming particle shapes. It is noteworthy that these transformations-- whether they are first-order or second-order-- can be experimentally realized in systems of reconfigurable particles that change geometry as a result of external stimuli such as change in ionic strength, pH, temperature, or pressure~\cite{GuoJPCC2007,LahannSoftMatter2009,TrungACSNano2010,GangACSNano2011}.

An uncertainty that plagues all numerical studies of packing is the inability to explore dense packings with large unit cells. This is because of the limitations of the utilized search algorithms that  become very inefficient when the number of particles in the unit cell becomes large. In this spirit, we confined our search to packings with a few particles in the unit cell, but we cannot rule out the possibility that denser packings of tTBPs with larger unit cells might exist.

Except for the degenerate quasicrystal, none of the ordered phases formed in self-assembly simulations of tTBPs is degenerate to a structure formed by the truncated tetrahedron system. This is not surprising since a tTBP is not the union of two truncated tetrahedra with the same truncation, and the bulkiness around its equatorial triangles makes it geometrically distinct. The added bulkiness is relatively small for small truncations, which explains why the degenerate quasicrystal still forms in the tTBP system. For higher truncations, however, a tTBP is so different from a corresponding pair of truncated tetrahedra that none of the structures formed by truncated tetrahedra is feasible. Truncated TBPs also self-assemble into fewer ordered structures than truncated tetrahedra, a fact that may be ascribed to the higher geometrical anisotropy of tTBPs.

We explain the observation that only one of the eight packing types self-assembles from the disordered fluid by using the predictive framework proposed by Damasceno~\emph{et~al.}~\cite{PabloScience2012}.  Structural differences in the first coordination shell between the tTBP packings and their disordered fluids makes most of these packings less likely to assemble from the fluid than a structure with similar local packing to that of the fluid, since the latter transformation will involve lower energy barriers. A similar phenomenon is observed for ellipsoids. Oblate and prolate ellipsoids with intermediate aspect ratios have asphericities similar to those of the truncated tTBPs considered in this study, and-- like our packings-- are notoriously hard to assemble into their stable SM2 phase from the fluid~\cite{DonevStillingerPRL2004, schillingPRE2007}. Further work is needed to better understand the role of shape sphericity on self-assembly propensity.

\acknowledgments
This material is based upon work supported by the DOD/ASDRE under Award No. N00244-09-1-0062, the National Science Foundation Award No. CHE 0624807, and the U.S. Department of Energy, Office of Basic Energy Sciences, Division of Materials Sciences and Engineering under Award \#DE-FG02-02ER46000.  Any opinions, findings, and conclusions or recommendations expressed in this publication are those of the author(s) and do not necessarily reflect the views of the DOD/ASDRE.  M. E. acknowledges support from the Deutsche Forschungsgemeinschaft. A.H.-A. acknowledges support from the University of Michigan Rackham Predoctoral Fellowship program.

\appendix

\section{Packings of truncated triangular bipyramids with two particles per unit cell}
\label{appendix:tTBPanalytical}

In the Appendix, we present analytical expressions for the intersection constraints, lattice vectors, and the packing fraction of densest packings. In this first section each unit cell contains two tTBPs that are related by inversion. We denote those particles with $X$ and $-X$, respectively. The lattice vectors are denoted by $\textbf{a},\textbf{b}$ and $\textbf{c}$. The offset vector connecting the centroids of $X$ and $-X$ is denoted by $\textbf{d}$. As a result, the centroids of $X$ are given by $n_a\textbf{a}+n_b\textbf{b}+n_c\textbf{c}$ with $n_a+n_c+n_c=0\text{ mod }2$, and the centroids of $-X$ are given by $n_a\textbf{a}+n_b\textbf{b}+n_c\textbf{c}+\textbf{d}$ with $n_a+n_c+n_c=1\text{ mod }2$. The volume of the unit cell is given by
\begin{eqnarray}
V &=& \det[\begin{array}{lll}\textbf{b}+\textbf{c},&\textbf{c}+\textbf{a},& \textbf{a}+\textbf{b}\end{array}].
\end{eqnarray}
The packing parameters, as denoted by $q_i$ in Section~\ref{subsection:packing_problem}, are therefore the lattice vectors $\textbf{a},\textbf{b}$ and $\textbf{c}$ and the offset $\textbf{d}$. We formulate the intersection constraints for each packing and provide lattice and offset vectors for the solution of Eq.~(\ref{eq:packing_opt}) consistent with these constraints. The maximum packing fractions are given in the main text and thus are not repeated here. The notation used in the present work is similar to the notation used in Ref.~\cite{Chen2010}.

\subsection{The packing P$_{2}1$}
There are a total of eleven intersection equations:

\begin{subequations}
	\footnotesize
	\label{eq:P$_{2}1$:intersections}
	\begin{align}
		E[\textbf{o},\textbf{q}]\cap (E[\textbf{s},\textbf{r}]+\textbf{a}+\textbf{b}) &\neq \emptyset,\displaybreak[0]\\
		E[\textbf{o},\textbf{r}] \cap (E[\textbf{s},\textbf{p}]+\textbf{b}+\textbf{c}) &\neq \emptyset,\displaybreak[0]\\
		V[\textbf{rs}] \cap (F[\textbf{o},\textbf{p},\textbf{q}]+\textbf{c}-\textbf{a}) &\neq \emptyset,\displaybreak[0]\\
		F[\textbf{o},\textbf{p},\textbf{q}]\cap (\textbf{d}-F[\textbf{o},\textbf{p},\textbf{q}]+\textbf{a}) &\neq \emptyset,\displaybreak[0]\\
		F[\textbf{o},\textbf{q},\textbf{r}]\cap(\textbf{d}-F[\textbf{o},\textbf{q},\textbf{r}]+\textbf{b}) &\neq \emptyset,\displaybreak[0]\\
		F[\textbf{o},\textbf{r},\textbf{p}] \cap (\textbf{d}-F[\textbf{o},\textbf{r},\textbf{p}]+\textbf{c}) &\neq \emptyset,\displaybreak[0]\\
		F[\textbf{s},\textbf{q},\textbf{r}]\cap(\textbf{d}-F[\textbf{s},\textbf{q},\textbf{r}]-\textbf{a}) &\neq \emptyset,\displaybreak[0]\\
		F[\textbf{s},\textbf{r},\textbf{p}]\cap(\textbf{d}-F[\textbf{s},\textbf{r},\textbf{p}]-\textbf{b}) &\neq \emptyset,\displaybreak[0]\\
		F[\textbf{s},\textbf{p},\textbf{q}]\cap(\textbf{d}-F[\textbf{s},\textbf{p},\textbf{q}]-\textbf{c}) &\neq \emptyset,\displaybreak[0]\\
		F[\textbf{o},\textbf{r},\textbf{p}]\cap(\textbf{d}-F[\textbf{o},\textbf{r},\textbf{p}]+\textbf{a}+\textbf{b}+\textbf{c}) &\neq \emptyset,\displaybreak[0]\\
		F[\textbf{s},\textbf{p},\textbf{q}]\cap (\textbf{d}-F[\textbf{s},\textbf{p},\textbf{q}]-\textbf{a}-\textbf{b}-\textbf{c}) &\neq \emptyset.
	\end{align}
\end{subequations}
The lattice vectors and the offset are given by
\begin{subequations}
	\begin{align}
\textbf{a} &= \tfrac{1}{320}(870+10t,321-25t,-21+45t) ,\displaybreak[0]\\
\textbf{b} &= \tfrac{1}{320}(-102-10t,831+25t,405-45t) ,\displaybreak[0]\\
\textbf{c} &= \tfrac{1}{320}(282+150t,-249+145t,741+35t) ,\displaybreak[0]\\
\textbf{d} &= \tfrac{1}{320}(38+10t,5+35t,-25-15t).
	\end{align}
\end{subequations}
The maximum packing fraction is given by Eq.~(\ref{eq:max_pf_P$_{2}1$}). This packing is valid for $0\le t\le t_1\approx0.2544$. $t_1$ is a root of the cubic equation $32000t_1^3-158825t_1^2+66310t_1-7117=0$.

\subsection{The packing P$_{2}2$}
There are a total of ten intersection equations:

\begin{subequations}
	\footnotesize
	\label{eq:P$_{2}2$:intersection}
	\begin{align}
		E[\textbf{o},\textbf{r}] \cap (E[\textbf{s},\textbf{p}]+\textbf{b}+\textbf{c} ) &\neq \emptyset,\displaybreak[0]\\
		E[\textbf{ro},\textbf{rs}] \cap (E[\textbf{p},\textbf{q}]+\textbf{c}-\textbf{a}) & \neq \emptyset,\displaybreak[0]\\
		F[\textbf{o},\textbf{p},\textbf{q}]\cap (\textbf{d}-F[\textbf{o},\textbf{p},\textbf{q}]+\textbf{a}) &\neq \emptyset,\displaybreak[0]\\
		F[\textbf{o},\textbf{q},\textbf{r}]\cap (\textbf{d}-F[\textbf{o},\textbf{q},\textbf{r}]+\textbf{b}) &\neq \emptyset,\displaybreak[0]\\
		F[\textbf{o},\textbf{r},\textbf{p}]\cap (\textbf{d}-F[\textbf{o},\textbf{r},\textbf{p}]+\textbf{c}) &\neq \emptyset,\displaybreak[0]\\
		F[\textbf{s},\textbf{q},\textbf{r}]\cap (\textbf{d}-F[\textbf{s},\textbf{q},\textbf{r}]-\textbf{a}) &\neq \emptyset,\displaybreak[0]\\
		F[\textbf{s},\textbf{r},\textbf{p}]\cap(\textbf{d}-F[\textbf{s},\textbf{r},\textbf{p}]-\textbf{b}) &\neq \emptyset,\displaybreak[0]\\
		F[\textbf{s},\textbf{p},\textbf{q}]\cap(\textbf{d}-F[\textbf{s},\textbf{p},\textbf{q}]-\textbf{c}) &\neq \emptyset,\displaybreak[0]\\
		F[\textbf{op},\textbf{oq},\textbf{or}]\cap(\textbf{d}-F[\textbf{op},\textbf{oq},\textbf{or}]+\textbf{a}+\textbf{b}+\textbf{c}) &\neq \emptyset,\displaybreak[0]\\
		F[\textbf{sp},\textbf{sq},\textbf{sr}]\cap(\textbf{d}-F[\textbf{sp},\textbf{sq},\textbf{sr}]-\textbf{a}-\textbf{b}-\textbf{c})&\neq \emptyset.
	 \end{align}
\end{subequations}
The lattice vectors and the offset are given by
\begin{subequations}
	\begin{align}
		\textbf{a} &= \tfrac1{36}(88+19t,67-92t,1-23t) ,\displaybreak[0]\\
		\textbf{b} &= \tfrac1{36}(16-101t,103-32t,37+t) ,\displaybreak[0]\\
		\textbf{c} &= \tfrac1{36}(40+19t,-17+16t,97-23t) ,\displaybreak[0]\\
		\textbf{d} &= \tfrac1{36}(-10+35t,5-10t,5-25t).
	\end{align}
\end{subequations}
The maximum packing fraction is given by Eq.~(\ref{eq:max_pf_P$_{2}2$}). This packing is valid for $0.2544\approx t_1\le t\le t_2=\frac13$.

\subsection{The packing P$_{2}3$}
There are a total of ten intersection equations:

\begin{subequations}
	\footnotesize
	\label{eq:P$_{2}3$:intersection}
	\begin{align}
		E[\textbf{o},\textbf{p}]\cap(E[\textbf{s},\textbf{q}]+\textbf{c}+\textbf{a}) &\neq \emptyset,\displaybreak[0]\\
		E[\textbf{o},\textbf{q}]\cap(E[\textbf{s},\textbf{r}]+\textbf{a}+\textbf{b}) &\neq \emptyset,\displaybreak[0]\\
		F[\textbf{o},\textbf{p},\textbf{q}]\cap(\textbf{d}-F[\textbf{o},\textbf{p},\textbf{q}]+\textbf{a}) &\neq \emptyset,\displaybreak[0]\\
		F[\textbf{o},\textbf{q},\textbf{r}]\cap(\textbf{d}-F[\textbf{o},\textbf{q},\textbf{r}]+\textbf{b}) &\neq \emptyset,\displaybreak[0]\\
		F[\textbf{o},\textbf{r},\textbf{p}]\cap(\textbf{d}-F[\textbf{o},\textbf{r},\textbf{p}]+\textbf{c}) &\neq \emptyset,\displaybreak[0]\\
		F[\textbf{s},\textbf{q},\textbf{r}]\cap(\textbf{d}-F[\textbf{s},\textbf{q},\textbf{r}]-\textbf{a}) &\neq \emptyset,\displaybreak[0]\\
		F[\textbf{s},\textbf{r},\textbf{p}]\cap(\textbf{d}-F[\textbf{s},\textbf{r},\textbf{p}]-\textbf{b}) &\neq \emptyset,\displaybreak[0]\\
		F[\textbf{s},\textbf{p},\textbf{q}]\cap(\textbf{d}-F[\textbf{s},\textbf{p},\textbf{q}]-\textbf{c}) &\neq \emptyset,\displaybreak[0]\\
		F[\textbf{op},\textbf{oq},\textbf{or}]\cap(\textbf{d}-F[\textbf{op},\textbf{oq},\textbf{or}]+\textbf{a}+\textbf{b}+\textbf{c})&\neq \emptyset,\displaybreak[0]\\
		F[\textbf{sp},\textbf{sq},\textbf{sr}]\cap(\textbf{d}-F[\textbf{sp},\textbf{sq},\textbf{sr}]-\textbf{a}-\textbf{b}-\textbf{c}) &\neq \emptyset.
	\end{align}
\end{subequations}
The lattice vectors and the offset are given by
\begin{subequations}
	\begin{align}
		\textbf{a} &= \tfrac{1}{36}(97-14t,40-8t,-17-2t) ,\displaybreak[0]\\
		\textbf{b} & =\tfrac{1}{36}(1-38t,88+4t,67-62t) ,\displaybreak[0]\\
		\textbf{c} & = \tfrac{1}{36}(37-2t,16-68t,103-26t) ,\displaybreak[0]\\
		\textbf{d} & = \tfrac{1}{36}(5-10t,-10+20t,5-10t).
	\end{align}
\end{subequations}
The maximum packing fraction is given by Eq.~(\ref{eq:max_pf_P$_{2}3$}). This packing is only valid for $\tfrac13=t_2\le t\le t_3\approx0.5010$. There is no analytical formula for $t_3$ because there is no analytical formula for $\phi_{4,1}(t)$.

\subsection{The packing P$_{2}4$}
There are a total of nine intersection equations:

\begin{subequations}
	\footnotesize
	\label{eq:P$_{2}4$:intersection}
	\begin{align}
		E[\textbf{op},\textbf{oq}]\cap(E[\textbf{sr},\textbf{sp}]+\textbf{c}+\textbf{a}) &\neq \emptyset,\displaybreak[0]\\
		E[\textbf{qo},\textbf{qr}]\cap(E[\textbf{ps},\textbf{pr}]+\textbf{b}-\textbf{c}) &\neq \emptyset,\displaybreak[0]\\
		E[\textbf{ro},\textbf{rp}]\cap(E[\textbf{qs},\textbf{qp}]+\textbf{c}-\textbf{a}) &\neq \emptyset,\displaybreak[0]\\
		E[\textbf{po},\textbf{pq}]\cap(E[\textbf{rs},\textbf{rq}]+\textbf{a}-\textbf{b}) &\neq \emptyset,\displaybreak[0]\\
		F[\textbf{o},\textbf{p},\textbf{q}]\cap(\textbf{d}-F[\textbf{o},\textbf{p},\textbf{q}]+\textbf{a})&\neq \emptyset,\displaybreak[0]\\
		F[\textbf{o},\textbf{q},\textbf{r}]\cap(\textbf{d}-F[\textbf{o},\textbf{q},\textbf{r}]+\textbf{b})&\neq \emptyset,\displaybreak[0]\\
		F[\textbf{o},\textbf{r},\textbf{p}]\cap(\textbf{d}-F[\textbf{o},\textbf{r},\textbf{p}]+\textbf{c})&\neq \emptyset,\displaybreak[0]\\
		F[\textbf{s},\textbf{p},\textbf{q}]\cap(\textbf{d}-F[\textbf{s},\textbf{p},\textbf{q}]-\textbf{c}) &\neq \emptyset,\displaybreak[0]\\
		F[\textbf{s},\textbf{q},\textbf{r}]\cap(\textbf{d}-F[\textbf{s},\textbf{q},\textbf{r}]-\textbf{a}+\textbf{b}-\textbf{c}) &\neq \emptyset.
	\end{align}
\end{subequations}
The lattice vector and the offset are given by
\begin{subequations}
	\begin{align}
		\textbf{a} &= \tfrac14(12-10t,24-22t,6-7t) ,\displaybreak[0]\\
		\textbf{b} &= \tfrac14(-4t,24-22t,18-13t) ,\displaybreak[0]\\
		\textbf{c} &= \tfrac14(8t,2t,6+5t) ,\displaybreak[0]\\
		\textbf{d} &= \tfrac14(-2+7t,-20+28t,-8+10t).
	\end{align}
\end{subequations}
The maximum packing fraction is given by Eq.~(\ref{eq:max_pf_P$_{2}4$}). This packing is valid for $0.5777\approx t_5\le t\le t_6=\frac23$. There is no analytical solution for $t_5$ because there is no analytical formula for $\phi_{4,2}(t)$.

\subsection{The packing P$_{2}5$\label{appendix:P2_5}}
There are a total of ten intersection equations:

\begin{subequations}
	\footnotesize
	\label{eq:P$_{2}5$:intersection}
	\begin{align}
		E[\textbf{op},\textbf{oq}]\cap(E[\textbf{sr},\textbf{sp}]+\textbf{a}+\textbf{b}) &\neq \emptyset,\displaybreak[0]\\
		E[\textbf{qo},\textbf{qr}]\cap(E[\textbf{ps},\textbf{pr}]+\textbf{b}-\textbf{c}) &\neq \emptyset,\displaybreak[0]\\
		F[\textbf{o},\textbf{p},\textbf{q}]\cap(\textbf{d}-F[\textbf{o},\textbf{p},\textbf{q}]+\textbf{a}) &\neq \emptyset,\displaybreak[0]\\
		F[\textbf{o},\textbf{q},\textbf{r}]\cap(\textbf{d}-F[\textbf{o},\textbf{q},\textbf{r}]+\textbf{b}) &\neq \emptyset,\displaybreak[0]\\
		F[\textbf{o},\textbf{r},\textbf{p}]\cap(\textbf{d}-F[\textbf{o},\textbf{r},\textbf{p}]+\textbf{c}) &\neq \emptyset,\displaybreak[0]\\
		F[\textbf{s},\textbf{q},\textbf{r}]\cap(\textbf{d}-F[\textbf{s},\textbf{q},\textbf{r}]-\textbf{a}) &\neq \emptyset,\displaybreak[0]\\
		F[\textbf{s},\textbf{r},\textbf{p}]\cap(\textbf{d}-F[\textbf{s},\textbf{r},\textbf{p}]-\textbf{a}-\textbf{b}+\textbf{c}) &\neq \emptyset,\displaybreak[0]\\
		F[\textbf{s},\textbf{p},\textbf{q}]\cap(\textbf{d}-F[\textbf{s},\textbf{p},\textbf{q}]-\textbf{b})&\neq \emptyset,\displaybreak[0]\\
		F[\textbf{qo},\textbf{qs},\textbf{qp}]\cap(\textbf{d}-F[\textbf{qo},\textbf{qs},\textbf{qp}]-\textbf{c}]) &\neq \emptyset,\displaybreak[0]\\
		F[\textbf{qo},\textbf{qs},\textbf{qp}]\cap(\textbf{d}-F[\textbf{qo},\textbf{qs},\textbf{qp}]+\textbf{a}+\textbf{b}-\textbf{c}) &\neq \emptyset.
	\end{align}
\end{subequations}
The lattice vectors and the offset are given by
\begin{subequations}
	\begin{align}
		\textbf{a} &= \tfrac1{56}(130+t,88-20t,118-149t) ,\displaybreak[0]\\
		\textbf{b} &= \tfrac1{56}(94-113t,136-92t,106+37t) ,\displaybreak[0]\\
		\textbf{c} &= \tfrac1{56}(250-179t,-44+10t,130+t) ,\displaybreak[0]\\
		\textbf{d} &= \tfrac1{56}(-38+29t,100-94t,-62+65t).
	\end{align}
\end{subequations}
The maximum packing fraction is given by Eq.~(\ref{eq:max_pf_P$_{2}5$}). This packing is valid for $\tfrac23=t_{6}\le t\le t_7=3-\tfrac15{\sqrt{105}}\approx0.9506$. $t_7$ is a root of the quadratic equation $5t_7^2-30t_7+24=0$.

\subsection{The packing P$_{2}6$}
There are a total of nine intersection equations:

\begin{subequations}
\footnotesize
	\label{eq:P$_{2}6$:intersection}
	\begin{align}
		E[\textbf{po},\textbf{pq}]\cap(E[\textbf{rs},\textbf{rq}]+\textbf{a}-\textbf{b}) &\neq \emptyset,\displaybreak[0]\\
		E[\textbf{qo},\textbf{qr}]\cap(E[\textbf{ps},\textbf{pr}]+\textbf{b}-\textbf{c})&\neq \emptyset,\displaybreak[0]\\
		E[\textbf{rs},\textbf{rp}]\cap(E[\textbf{qo},\textbf{qp}]+\textbf{c}-\textbf{a})&\neq \emptyset,\displaybreak[0]\\
		F[\textbf{o},\textbf{q},\textbf{r}]\cap(\textbf{d}-F[\textbf{o},\textbf{q},\textbf{r}]+\textbf{b})&\neq \emptyset,\displaybreak[0]\\
		F[\textbf{o},\textbf{r},\textbf{p}]\cap(\textbf{d}-F[\textbf{o},\textbf{r},\textbf{p}]+\textbf{c}) &\neq \emptyset,\displaybreak[0]\\
		F[\textbf{s},\textbf{p},\textbf{q}]\cap(\textbf{d}-F[\textbf{s},\textbf{p},\textbf{q}]-\textbf{a}) &\neq \emptyset,\displaybreak[0]\\
		F[\textbf{s},\textbf{q},\textbf{r}]\cap(\textbf{d}-F[\textbf{s},\textbf{q},\textbf{r}]+\textbf{b}-2\textbf{a}) &\neq \emptyset,\displaybreak[0]\\
		F[\textbf{op},\textbf{oq},\textbf{or}]\cap(\textbf{d}-F[\textbf{op},\textbf{oq},\textbf{or}]+\textbf{a})&\neq \emptyset,\displaybreak[0]\\
		F[\textbf{sp},\textbf{sq},\textbf{sr}]\cap(\textbf{d}-F[\textbf{sp},\textbf{sq},\textbf{sr}]+\textbf{c}-2\textbf{a})&\neq \emptyset.
	\end{align}
\end{subequations}
The lattice vectors and the offset vector are given by
\begin{subequations}
	\begin{align}
		\textbf{a} &= \tfrac{1}{40}(88-20t,118-65t,58+25t) ,\displaybreak[0]\\
		\textbf{b} &= \tfrac{1}{40}(-104+100t,46-5t,106+25t) ,\displaybreak[0]\\
		\textbf{c} &= \tfrac{1}{40}(-56+100t,-146+115t,34+85t) ,\displaybreak[0]\\
		\textbf{d} &= \tfrac{1}{40}(156-130t,90-55t,-30+5t).
	\end{align}
\end{subequations}
The maximum packing fraction is given by Eq.~(\ref{eq:max_pf_P$_{2}6$}). This packing is valid for $0.9506\approx t_7\le t\le t_8=1$.

\section{Packings of truncated triangular bipyramids with four particles per unit cell}\label{appendix:4pPackings}
For the two four-particle packings P$_{4}1$ and P$_{4}2$ the polyhedron $X$ as well as the lattice vectors $\textbf{a}$, $\textbf{b}$, and $\textbf{c}$ are as defined in Appendix~\ref{appendix:tTBPanalytical}. The other three particles are given by $-X$ and $\pm MX$ with the transformation matrix $M$ given by
\begin{eqnarray}
	M & = & \left[
		\begin{array}{ccc}
			1-2x^2	& -2xy & -2xz\\
			-2yx & 1-2y^2 & -2yz\\
			-2zx & -2zy & 1-2z^2
		\end{array}
	\right].
\end{eqnarray}
Here $(x,y,z)=(a_x,a_y,a_z)/\|\textbf{a}\|$ is a unit vector in the direction of $\textbf{a}$. $M$ describes a reflection across a plane that passes through the origin and is perpendicular to $\textbf{a}$.  The offsets between $X$, $-X$, and $\pm MX$ are given by $\textbf{d}, \textbf{e}$, and $\textbf{f}$, respectively. The optimization problem is then solved with the intersection equations for the particular packing alongside the constraints
\begin{subequations}
\label{eq:appendix:symm_const}
	\begin{align}
		M\textbf{a} &=-\textbf{a},\quad&
		M\textbf{b} &= \textbf{b}-\textbf{a},\displaybreak[0]\\
		M\textbf{c} &= \textbf{c},\quad&
		M\textbf{d} &= \textbf{f}-\textbf{e},\displaybreak[0]\\
		\textbf{a}\cdot(2\textbf{b}-\textbf{a}) &= 0,\quad&
		\textbf{a}\cdot\textbf{c} &= 0,\displaybreak[0]\\
		\textbf{a}\cdot(\textbf{e}-\textbf{d}) &= 0,\quad&
		\textbf{a}\cdot\textbf{f} &= 0.
	\end{align}
\end{subequations}
For the packing P$_{4}1$ there are a total of ten intersection equations:
\begin{subequations}
%\footnotesize
	\begin{align}
		F[\textbf{o},\textbf{q},\textbf{r}]\cap(V[\textbf{ps}]+\textbf{a}) &\neq \emptyset\displaybreak[0]\\
		E[\textbf{oq},\textbf{or}]\cap(E[\textbf{sp},\textbf{sq}]+\textbf{b}) &\neq \emptyset\displaybreak[0]\\
		F[\textbf{o},\textbf{r},\textbf{p}]\cap(\textbf{d}-F[\textbf{o},\textbf{r},\textbf{p}]) &\neq \emptyset\displaybreak[0]\\
		F[\textbf{s},\textbf{q},\textbf{r}]\cap(\textbf{d}-F[\textbf{s},\textbf{q},\textbf{r}]+\textbf{a}-\textbf{b})&\neq \emptyset		\displaybreak[0]\\				
		F[\textbf{s},\textbf{r},\textbf{p}]\cap(\textbf{d}-F[\textbf{s},\textbf{r},\textbf{p}]-\textbf{b})&\neq \emptyset\displaybreak[0]\\
		V[\textbf{qo}]\cap(\textbf{e}+MF[\textbf{o},\textbf{q},\textbf{r}]+\textbf{a})&\neq \emptyset\displaybreak[0]\\
		F[\textbf{o},\textbf{q},\textbf{r}]\cap(\textbf{e}+MV[\textbf{qo}]+\textbf{a}+\textbf{c})&\neq \emptyset\displaybreak[0]\\
		F[\textbf{s},\textbf{p},\textbf{q}]\cap(\textbf{e}+ME[\textbf{ro},\textbf{rp}])&\neq \emptyset\displaybreak[0]\\
		E[\textbf{ro},\textbf{rp}]\cap(\textbf{e}+MF[\textbf{s},\textbf{p},\textbf{q}]+\textbf{c})&\neq \emptyset\displaybreak[0]\\
		E[\textbf{o},\textbf{q}]\cap(\textbf{f}-ME[\textbf{o},\textbf{q}]) &\neq \emptyset
	\end{align}
\end{subequations}
The same intersection equations are satisfied in the packing P$_{4}2$. Additional there are the following two intersections:
\begin{subequations}
\footnotesize
	\begin{align}
		V[\textbf{qs}]\cap(\textbf{f}-MF[\textbf{s},\textbf{p},\textbf{q}]+\textbf{a}-\textbf{b})&\neq \emptyset\displaybreak[0]\\
		F[\textbf{s},\textbf{p},\textbf{q}]\cap(\textbf{f}-MV[\textbf{qs}]-\textbf{b})&\neq \emptyset
	\end{align}
\end{subequations}
Because of the nontrivial form of $M$, Eq.~(\ref{eq:packing_opt}) cannot be solved analytically for four-particle packings. All solutions are obtained numerically. The packing P$_{4}1$ is valid for $0.5010\approx t_3\le t\le t_4\approx 0.5321$ while P$_{4}2$ is valid for $0.5321\approx t_4\le t\le t_5\approx 0.5776$. Note that there is no analytical solution for $t_3,t_4$ and $t_5$ and there is no analytical formula for $\phi_{4,1}(t)$ and $\phi_{4,2}(t)$.

\bibliographystyle{apsrev}
\bibliography{References}

\end{document}